\documentclass[preprint, aps, pre, eqsecnum,amsmath,amssymb,showpacs]{revtex4}              

 \usepackage[final]{graphicx}  

 \newcommand{\Nabla}{\mbox{\bf\boldmath $\nabla$}}                              
                                                              
 \renewcommand{\vec}[1]{{\bf #1}}

\begin{document}

 \title{\bf 
  Traveling Wave Fronts and Localized Traveling Wave Convection in Binary Fluid 
  Mixtures}                                

 \author{D.~Jung and M.~L\"ucke}                                   
 \affiliation{Institut f\"ur Theoretische Physik, Universit\"at des Saarlandes,      
 Postfach 151150, \\ D-66041 Saarbr\"ucken, Germany}                              
   
 \date{\today}                                                                   

 \begin{abstract}
  Nonlinear fronts between spatially extended traveling wave convection (TW) 
  and quiescent fluid and spatially localized traveling waves (LTWs) are 
  investigated in quantitative detail in the bistable regime of binary fluid 
  mixtures heated from below. 
  A finite-difference method is used to solve the full hydrodynamic field 
  equations in a vertical cross section of the layer perpendicular
  to the convection roll axes. Results are presented for ethanol-water
  parameters with several strongly negative separation ratios where TW solutions
  bifurcate subcritically. Fronts and LTWs
  are compared with each other and similarities and differences are elucidated.
  Phase propagation out of the
  quiescent fluid into the convective structure entails a unique selection of
  the latter while fronts and interfaces where the phase moves into the
  quiescent state behave differently. Interpretations of various experimental
  observations are suggested. 
 \end{abstract}

 \maketitle                                                                  
 \vskip2pc

 
\section{INTRODUCTION}

Many nonlinear dissipative systems that are driven sufficiently far away from 
 thermal equilibrium show selforganization out of an unstructured state: A 
 structured one can appear that is characterized by a (spatially extended) 
pattern which retains some of the symmetries of the system \cite{CH93}.
 Convection in binary miscible fluids like ethanol-water, $^3$He 
$-^4$He, or various gas mixtures is an example of such systems. It shows rich 
and interesting pattern formation behavior and it is paradigmatic for problems 
related to instabilities, bifurcations, and
selforganization with complex spatiotemporal behavior.

Compared to convection in one-component 
fluids like, e.g., pure water the spatiotemporal properties are far more 
complex. The reason is that concentration variations 
which are generated via thermodiffusion, i.e., the Soret effect by 
externally imposed and by internal temperature gradients influence the buoyancy,
i.e., the driving force for convective flow. The latter in turn mixes by
advectively redistributing concentration. This nonlinear advection
gets in developed convective flow typically much larger than the smoothening 
by linear diffusion --- P\'eclet numbers measuring the strength
of advective concentration transport relative to diffusion are easily 
$\cal O$(1000). Thus, the concentration balance is strongly nonlinear 
giving rise to strong variations of the concentration field and to
boundary layer behavior. In contrast to that, momentum and 
heat balances remain weakly nonlinear close to onset as in pure fluids implying
only smooth and basically harmonic variations of velocity and temperature fields
as of the critical modes. 

Without the thermodiffusive Soret coupling between temperature and
concentration any initial concentration deviation from the mean diffuses away 
and influences no longer the balances of the other fields. 
Hence, the feedback interplay between ({\em i}) the Soret generated 
concentration variations, ({\em ii}) the resulting modified buoyancy, and
({\em iii}) the 
strongly nonlinear advective transport and mixing causes binary mixture 
convection to be rather complex with respect to its spatiotemporal
properties and its bifurcation behavior.
Take for example the case of negative Soret coupling, $\psi < 0$, 
between temperature and concentration fields \cite{coupling} when the lighter
component migrates to the colder regions thereby stabilizing the density 
stratification in the quiescent, laterally homogeneous conductive fluid state.
Then the above 
described feedback interplay generates oscillations. In fact the buoyancy 
difference in regions
with different concentrations was identified already in \cite{WKPS85} as the 
cause for traveling wave convection.

Oscillatory convection appears in the form of the
transient growth of convection at supercritical heating, in spatially
extended  
nonlinear traveling wave (TW) and standing wave solutions that branch 
in general subcritically
out of the conductive state via a common Hopf bifurcation, in spatially
localized traveling wave (LTW) states, and in various types of fronts. TW and LTW convection 
has been studied experimentally and theoretically  for some time
\cite{CH93,Moses87Heinrichs87,Behringer9091,Winkler92,Kolodner94,Platten96,Kaplan94,Surko91,
      Aegert01,Ning97b,Batiste01,Jung02}.
The transient oscillatory growth of convection was investigated by
numerical simulations \cite{FL}. Nonlinear standing wave solutions were obtained 
only recently \cite{MJL04,JML04}. Freely propagating convection fronts that
connect subcritically bifurcating nonlinear TW convection with the {\em stable}
quiescent fluid do not seem to have been investigated in detail beyond some
first preliminary results \cite{Bensimon90,Kolodner92Fronts,Bu99}. Here we determine such fronts in
quantitative detail and compare their properties with those of LTWs.

 In narrow rectangular and annular channels convection occurs in 
the form of rolls with axes oriented perpendicular to the long sidewalls 
\cite{Surko87,CH93}. These stuctures can efficiently be described in the two
dimensional vertical $x-z$ cross section in the middle of the channel 
perpendicular to the roll axes ignoring variations in axis direction. 
Furthermore, these convection structures have relevant phase gradients only in  
$x$-direction thus causing effectively one dimensional patterns \cite{AlBa04}. 

When comparing 
experiments with analytical calculations or numerical simulations
performed under the above described conditions it is useful to do that on the
basis of reduced Rayleigh numbers, $r=R/R^0_c$, with $R^0_c$ being the critical
one for onset of pure fluid convection for the respective experiment, analytical
method, or numerical method. This significantly reduces the dependence 
of, say, the bifurcation diagrams of convective states on the specific
geometry of the respective set-up. In laterally unbounded systems the analytical
value for $R^0_c$ is 1707.762.

{\em Localized traveling waves.} \hspace{0.3cm}
For weak negative Soret coupling one has observed in experiments a competition 
between homogeneous laterally extended TW convection and so-called dispersive 
chaos with an irregular repetitive formation and collapse of spatially localized
TW pulses \cite{Kolodner90Glazier91,Kaplan94}. During the pulse formation their 
drift velocities can drop abruptly 
to about a tenth of the initial group velocity \cite{Kolodner90Glazier91}.
We consider this to be a characteristic signal that the lateral redistribution
of concentration over the pulse \cite{Jung02} becomes important and that the 
strongly nonlinear dynamics sets in. For more negative $\psi \lesssim -0.06$ the 
collapse is in general less dramatic. There, convection is dominated 
by isolated strongly peaked localized states. Eventually, at $\psi\simeq -0.07$ 
a regime is reached where stable LTWs coexist near onset with extended 
TWs \cite{Barten95I,Barten95II,Luecke98,Moses87Heinrichs87,Kolodner88,
Kolodner91a,Kolodner94}.
Increasing the Soret coupling strength further to more negative $\psi$  
the band ($r^{LTW}_{min},r^{LTW}_{max}$) of Rayleigh numbers in which stable 
LTWs exist increases monotonically while shifting upwards as a whole ---  
$r^{LTW}_{max}(\psi)$ grows stronger than $r^{LTW}_{min}(\psi)$. Simultaneosly,
the lower band limit for the existence of extended TW states 
$r^{TW}_{min}(\psi)$, i.e., the lowest saddle-node of TWs increases even steeper
so that eventually for $\psi \lesssim -0.4$ the complete LTW band comes to lie 
below the existence range of TWs, $r^{LTW}_{max} \leq r^{TW}_{min}$
\cite{Jung02}.

LTWs consist of slowly drifting, spatially confined convective regions that are 
embedded in the quiescent fluid.
These intriguing structures have been investigated in experiments
\cite{Moses87Heinrichs87,Kolodner88,Bensimon90,Niemela90,Behringer9091,
Kolodner9091,Steinberg91,Kolodner91a,
Surko91,Kolodner91c,Kolodner91d,Kolodner93II,Kolodner94} and numerical simulations
\cite{Barten91,Barten95II,Luecke98,Jung02}. A discussion of
various theoretical models aiming at their explanation is contained in 
Sec.~\ref{SEC:LTWmodels}. Roll vortices grow in a LTW structure out
of the quiescent fluid at one end, travel with spatially varying phase velocity 
$v_p(x)$ to the other end, and decay there back into the basic state. The two 
interfaces to conduction and with it the whole convective region move with 
constant, uniquely selected drift velocity $v_d$. The latter is a 
function of $r, \psi$ with magnitude much smaller than the phase velocities.
Also the oscillation frequency of the LTW is uniquely selected; it is constant 
in space and time in the frame that is comoving with its drift velocity. 
And finally, the length $l(r,\psi)$ of the convective region of stable LTWs and 
their spatial stucture are uniquely selected. This length grows with increasing
heating $r$. 

A central role for the stable existence of LTWs plays a large-scale
mean concentration current. Extending over the whole LTW it redistributes
concentration and thereby changes the buoyancy in a decisive way \cite{Luecke98}.
This effect 
can sustain LTWs even at low $r$ where no extended TWs exist \cite{Jung02}.

{\em Blinking states in rectangular channels.} \hspace{0.3cm}
The LTW confinement of convection occurring in translationally invariant annular 
channels is obviously an inherent process of the hydrodynamic balances.
But one has also observed end-wall-assisted or at least end-wall-modified 
confinement of convection close to the ends of rectangular channels.
The weakly nonlinear varieties of such a confinement can largely be understood
in terms of the convective behavior
of TW packets, their reflection properties at the end walls, and the 
destructive interaction 
between left and right traveling patterns \cite{Cross868889,Brand86,Fineberg90}.
These effects give rise near onset to a wide range of weakly nonlinear and 
effectively low dimensional spatiotemporal behavior that depends sensitively
on the specific experimental set-up like, e.g., the end-wall boundary
conditions and the system length  
\cite{Kolodner8889,Fineberg88Steinberg89,Bestehorn89,Kolodner93,Batiste01}.
While the linear eigenmodes of such sytems ('linear counterpropagating 
waves' or 'chevrons') \cite{Kolodner86,Surko87,Kolodner87,Batiste01} 
are laterally symmetric or antisymmetric
localization sets in via a temporal amplitude modulation. Thereby
convection is alternatingly weakened and enhanced in the left and the right part
of the system part giving rise to a 'blinking' state  
\cite{Surko87,Kolodner8889,Fineberg88Steinberg89,Steinberg91,Kolodner93}.
The so-called 'chaotic blinking' states \cite{Kolodner8889,
Fineberg88Steinberg89,Steinberg91,Kolodner93} seem to be the 
analogue of the 'chaotic dispersive' pulse formation in annular containers 
\cite{Steinberg91,Kaplan94}. Also 'blinking' modes with different 
frequencies at
both ends of the channel were observed \cite{Fineberg88Steinberg89,Steinberg91}.
But their possible relation to a large-scale mean concentration variation 
\cite{Moses86} produced by nonlinear propagating waves in a finite cell has not 
been discussed.

{\em Wall-attached structures.} \hspace{0.3cm}
At larger $r$ one has observed wall-attached TW structures with amplitudes
confined to the vicinity of one or both end walls. These wall-attached 
convective patches 
\cite{Moses87Heinrichs87,Kolodner8889,Fineberg88Steinberg89,Kolodner90,Kolodner91b,Yahata91,
Ning96,Ning97a} are closely related to free LTWs \cite{Niemela90}. 
They are strongly nonlinear as indicated by their low frequency 
\cite{Kolodner8889,Fineberg88Steinberg89}.
Moreover, their spatial structure and their region of existence is largely
unaffected by the details
of the lateral boundaries or by the container length in contrast to the linear
and weakly nonlinear behavior described above \cite{Fineberg88Steinberg89}. The 
more extensive wall-attached structures show some similarities with front-like
states. Note, however, that here the source or the sink of the propagating rolls
is pinned near a wall and the interface to the quiescent fluid in the bulk of the
channel does not move \cite{Kolodner90}.

{\em Our numerical simulations.} \hspace{0.3cm}
Our numerical simulations have been performed in order to elucidate in
quantitative detail the properties of relaxed nonlinear TW convection structures
that contain an interface (or two of them) to the
quiescent fluid as an integrated structural element. We compare
for a wide range of Soret coupling strengths front states and LTW states
showing what they have in common and how they differ. We
focus our interest to those parameters where the quiescent conductive state of
the fluid is stable and where the solutions describing spatially extended,
laterally periodic TW convection bifurcate subcritically out of it.

The system we have in mind is a binary fluid layer of thickness $d$
which is bounded by two solid horizontal plates perpendicular to 
the gravitational acceleration $\vec{g}$.
The fluid might be a mixture of water with the lighter component ethanol at a
mean concentration $\overline{C}$.
It is heated from below. The temperatures at the plates are 
 $\overline{T} \pm \Delta T/2$.
The variation of the fluid density $\rho$ due to temperature and
concentration variations is governed by the linear thermal and solutal
expansion coefficients
$ \alpha = - \frac{1}{\rho}\frac{\partial\rho}{\partial {T}} $ and
$ \beta = - \frac{1}{\rho}\frac{\partial\rho}{\partial {C}} $, respectively.
Both are positive for ethanol-water.  The solutal diffusivity
of the binary  mixture is $D$, its thermal diffusivity is $\kappa$, and its
viscosity is $\nu$.
Length and time is scaled by $d$ and $d^2/\kappa$, respectively, so that
velocity is measured in units of $\kappa/d$.
Temperatures are reduced by the vertical temperature
difference $\Delta T$ across the layer and concentrations
by $\frac{\alpha}{\beta}\Delta T$.
The scale for the pressure is given by $\frac{\rho\kappa^2}{d^2}$.

Then, the balance equations for mass, momentum, heat, and concentration
  \cite{LL66,Platten84} read in Oberbeck--Boussinesq approximation \cite{Barten95I}
 \begin{subequations}
 \label{eq:baleqs}
 \begin{eqnarray}
 \Nabla \cdot \vec{u} = 0 \label{eq:baleqmass}\\
 \partial_t\, {\bf u}   =    - {\bf \mbox{\boldmath $\nabla$} }\,
({\bf u : u}  +  p
- \sigma \, {\bf \mbox{\boldmath $\nabla :  $}\,\, u }) +
{\bf B} \,\,\,;\,\,\,  
{\bf B}    =    \sigma\, R\, (\delta T + \delta C) {\bf e}_z \label{eq:baleqveloc}\\ 
 \partial_t\delta T  =  
  - \Nabla \cdot \left[ \vec{u}\delta T - \Nabla \delta T\right]\label{eq:baleqheat}\\
 \partial_t \delta C  = 
  - \Nabla \cdot \left[ \vec{u} \delta C - L\Nabla\left(\delta C -\psi \delta T\right) \right]\ .
  \label{eq:baleqconc}
 \end{eqnarray}
 \end{subequations}
Here, $\delta T$ and $\delta C$ denote deviations of
the temperature and concentration fields, respectively, from their 
mean $\overline{T}$ and $\overline{C}$ and $\bf B$ is the buoyancy.
The Dufour effect \cite{HLL92,HL95}
that provides a coupling of concentration gradients into
the heat current and a change of the thermal diffusivity
is discarded in (\ref{eq:baleqheat}) since it is relevant
only in few binary gas mixtures \cite{LA96} and possibly in
liquids near the liquid--vapor critical point \cite{LLT83}.

Besides the Rayleigh number $R=\frac{\alpha g d^3}{\nu \kappa}\Delta T$
measuring the thermal driving of the fluid three additional
numbers enter into the field equations: the Prandtl number
$\sigma=\nu/\kappa$, the Lewis number $L=D/\kappa$, and the separation ratio 
$\psi=-\frac{\beta}{\alpha}\frac{k_T}{\overline{T}}= 
- S_T\overline{C}(1-\overline{C})\frac{\beta}{\alpha}$. 
Here $k_T = \overline{T}\,\overline{C}(1-\overline{C})S_T$ is the thermodiffusion coefficient
\cite{LL66} and $S_T$ the Soret coefficient. They measure changes
of concentration fluctuations due to temperature gradients in the fluid.
$\psi$ characterizes the sign and the strength of the Soret effect. 
Negative Soret coupling $\psi$ (i.e., positive $S_T$ for mixtures like ethanol
water with positive $\alpha$ and $\beta$) induces concentration gradients of 
the lighter component  that are antiparallel to temperature gradients. 
In this situation, the buoyancy induced by solutal changes in density is opposed
to the thermal buoyancy.

When the gradient of the total buoyancy exceeds a threshold, convection sets in
 --- typically
in the form of straight rolls. For sufficiently negative $\psi$ the primary 
instability is oscillatory. Ignoring field
variations along the roll axes we describe here 2D convection in an
$x$--$z$ plane perpendicular to the roll axes with a velocity field
\begin{equation}
\vec{u}(x,z,t) = u(x,z,t)\,\vec{e}_x + w(x,z,t)\,\vec{e}_z \, .
\end{equation}

To find the time-dependent solutions of the above partial differential equations
subject to realistic horizontal boundary conditions \cite{Barten95I}
we performed numerical simulations with a modification of 
the SOLA code that is based on the MAC method \cite{MAC-SOLA,PT83}. 
This is a finite-difference method of second order in space formulated on 
staggered grids for the different fields. The Poisson equation for the pressure 
field that results from taking the divergence of (\ref{eq:baleqveloc}) was solved
iteratively with the artificial compressibility   
method \cite{PT83} by incorporating a multi-grid technique. 
  
Throughout this paper we consider mixtures with $L=0.01$, $\sigma=10$, 
and various negative values of $\psi$ that are easily accessible with 
ethanol-water experiments. The paper is organized as follows: In 
Sec.~\ref{Sec:Fronts} we first describe our methods for characterizing the
various convective states. Then we present results for the two different types of
TW front states that can arise in laterally homogeneous mirror symmetric systems
with either the phase propagating out of the quiescent fluid or into it. Also
transient two-front structures are discussed. Sec.~\ref{SEC:LTW} deals with LTW
states and their relation to fronts. The transient dynamics towards the selected
LTW, the stabilization via front repulsion, the difference between long and short
LTWs, and a critical appraisal of LTW models are topics covered here. In 
Sec.~\ref{SEC:Compare-exp} we present a comparison with experiments and a
discussion. The last section contains a conclusion.

 \section{Fronts} \label{Sec:Fronts} 

Here we discuss front solutions where part of the system is occupied by the 
quiescent fluid while the other one shows fully developed, saturated, strongly 
nonlinear TW convection with laterally homogeneous amplitude. 
Strictly speaking these two states are realized only in the two opposing limits
of $x\to \pm \infty$.
We focus our investigation of fronts on parameters where the quiescent fluid 
state is stable and where the TW solutions bifurcate subcritically out of the
conductive state. Then, any {\it linear} growing and spreading of 
infinitesimal, localized convective perturbations in the
quiescent fluid which could possibly dominate the low amplitude behavior 
of fronts as in the case of an unstable zero amplitude state 
\cite{vSaarloos9092,vSaarloos03} is absent.

Little general is known about pattern forming fronts in real bistable systems
\cite{CH93}. Most of the research activities were centered on fronts in the 
quintic Ginzburg-Landau equation \cite{vSaarloos9092,vHvSHo93,CoWaSMi99,CoCh01,
SzLu03,Rabaud03}. 
One can expect that the front properties are fixed by a strongly 
nonlinear eigenvalue problem 
describing a heteroclinic orbit between the two involved states.
Some of these front solutions will be unstable. There might be also  
multistable coexistence of fronts so that depending on initial conditions and 
on the history of the (control) parameters different fronts could finally be 
realized. We call a front uniquely selected when our numerical simulations 
indicated that different formation processes ended in the same front for a fixed
parameter combination.  

Fronts can be classified into coherent and incoherent ones \cite{vSaarloos03}.
We focus here on the first kind which in our system are characterized as 
strictly time periodic 
states in a frame that is comoving with the front's velocity $v_F$.  
Such a front state being monochromatic is a global nonlinear mode. Its 
frequency is an eigenvalue, i.e., a global constant in space and time so that 
the convection oscillations have everywhere the same period.

Thus, we do not consider here, e.g., complex large scale or chaotic 
spatiotemporal interface behavior. The coherent 
fronts of the various hydrodynamic fields and quantities in this paper have 
a smooth and basically monotonous profile which connects  
the quiescent fluid with the nonlinear saturated extended TW. The transition 
region between conduction and convection that is characterized by large 
amplitude variations is quite short and consists typically 
only of about 3 -4 convection rolls. We call this transition region also
the interface between conduction and convection.

If the selected front pattern is incompatible with any stable bulk structure 
there are two possibilities:
({\it i}) A perturbation in the unstable convection bulk grows and expands 
towards 
the interface. This would destroy front coherence and could lead to more 
complex large scale variations, perhaps chaotic spatiotemporal behavior.
({\it ii}) The interface region is only convectively unstable against
perturbations of the bulk nonlinear TW. Then initially localized
perturbations would be advected out of every finite system and could not  
reach the unperturbed interface region of the front.

 \subsection{Methods of characterization} \label{SEC:FRONT-charact}

 \subsubsection{Definitions} \label{SEC:FRONT-def}

We call a front to be of type $+$ when its envelope grows at $x=-\infty$ out 
of the basic quiescent state.
Otherwise it is a $-$front. Then the amplitude falls to zero at $x=+\infty$ 
\cite{Buechel00}. The phase of the convection pattern in a front state of type
$+$ can either propagate to the left or to the right and similarly for the 
$-$front state. Hence, one would have 
to discuss four front states separately. However, because of the invariance 
of the system under $x \to -x$  
a $+$front state with positive (negative) phase velocity $v_p$ is the mirror
image of the $-$front state with negative (positive) $v_p$. Therefore, it
suffices to consider only the front states that consist of roll vortices 
traveling, say, in positive $x$-direction and to use the superscript $+$ or
$-$ to identify the properties of the front in question in a unique way. So,
the phase velocities of all oscillatory convective structures investigated in 
this paper are positive. We call the direction of positive $x$ into which
the phase propagates also 'downstream' and the opposite one 'upstream'.

So, to sum up our notation: In a $+$front state the quiescent fluid is located
'upstream' and a source of phase with the latter propagating out of the
conductive state into convection. In a $-$front the quiescent fluid is located 
in 'downstream' direction and a sink since phase moves out of convection 
into conduction. 

Fig.~\ref{Fig:frontpics} shows fronts of each type. Under the $+$front (left
half of Fig.~\ref{Fig:frontpics}) convection rolls grow out of the
quiescent fluid and saturate in a 'downstream' bulk TW. On the other hand,
a $-$front (right half of Fig.~\ref{Fig:frontpics}) annihilates
roll vortices. In this process their phase velocity is accelerated [cf. the 
increase in the lateral profile of $v_p(x)$ in Fig.~\ref{Fig:frontpics}(f)] 
and they are stretched horizontally. 

It is clear from Fig.~\ref{Fig:frontpics} that the quiescent (convecting) region
expands into the convecting (quiescent) one when the velocity $v^+_F$ of the 
$+$front is positive (negative) and vice versa for the $-$front.

 \subsubsection{Mixing number}  \label{SEC:FRONT-mix}

In order to monitor how well the fluid is mixed along the front we always
determined for the relaxed front states the mixing number 
\begin{equation} \label{Eq:M(x)}
M(x)= 
\left[< \overline{(\delta C)^2} > /  \overline{ (\delta C_{cond})^2}\right]^{1/2}
\end{equation}
as a function of lateral position $x$. It basically
measures the mean square of the deviations $\delta C(x,z,t)$ of the concentration
field from its global mean: the overbars imply a vertical average and the
brackets a temporal average at the specific horizontal location $x$ in the
frame comoving with the front velocity $v_F$.
The subscript {\it cond} denotes the reference quiescent 
conductive state with its linear vertical concentation variation. The mixing
number is defined such that $M=0$ 
in a perfectly mixed fluid and $M=1$ in the quiescent state.

In laterally extended TWs $\omega$ and with it $v_p$ increase when the 
concentration variations become larger \cite{Barten95I,Luecke98}. In fact, there
is a
universal scaling relation between $M$ and $\omega$ \cite{HoBuLu97} which shows
that $M$ and $v_p$ are linearly related to each other.
This relation also holds for the bulk part of front states far away from the
interface where the convection is TW-like with only slow spatial amplitude 
variation (Fig.~\ref{Fig:frontpics}).

 \subsubsection{Concentration current}  \label{SEC:FRONT-current}

The phase shift between the concentration and velocity waves in the TW-like 
bulk of the front states sustains as in extended TW states a mean lateral
concentration current $<\vec{J}>(x,z)$ \cite{Linz88,Barten89,Barten95I,Luecke98}:
\begin{equation} \label{Eq:J}
<\vec{J}>= < \vec{u} \delta C - L \Nabla (\delta C - \psi \delta T) >
\end{equation}
where $\vec{u}$ is the velocity field and $\delta T$ the temperature deviation 
from the global mean. Again, the brackets imply a temporal average in the 
frame that is comoving with the front velocity. The Lewis number $L=0.01$ 
being rather small in our simulations
implies that $<\vec{J}>$ is dominated by the advective contribution except in
those boundary regions in which  $\vec{u}$ becomes small. 

The vertical variation of $<\vec{J}>$ is such that positive (negative)
$\delta C$ is transported in phase direction in the upper (lower) half of the 
layer. This transport causes a large-scale concentration redistribution in a 
front state between its TW bulk and its interface to the quiescent fluid and it 
is responsible for the different characteristic structures of the interfaces in
a $+$ and a $-$front as we will see further below.

 \subsubsection{Preparation and lateral boundary conditions}
 \label{SEC:FRONT-prep}

We simulated systems containing up to 160 rolls.
The initial state was prepared by filling one half of the sytem with a nonlinear 
TW that was previously generated with periodic boundaries to have some fixed 
wavelength $\lambda$. The other half contained the stable temperature and 
concentration distribution of the pure quiescent basic state.

To simulate $+$fronts in infinite systems that connect to developed TW convection 
with some wavelength $\lambda$ far away from the interface 
between conduction and convection we imposed at the 'downstream' boundary $x=L$ 
of our computation domain the periodicity condition $f(L)=f(L-\lambda)$. 
For the case of $-$fronts we found that imposing the analogous condition at the
'upstream' boundary of the developed TW part at $x=-L$ typically will introduce 
perturbations that can grow in 'downstream'
direction for example when the TW region is Eckhaus unstable.
The different aspects of the stability of $+$ and $-$fronts are discussed 
further below in the paper.
 
After a relaxation time of typically 100 to 200 
vertical thermal diffusion times 
we then could observe under certain conditions a coherent front state connecting 
a quiescent region of the system to a TW with asymptotic wavelength $\lambda$.
Here the fact that the frequency $\omega$ of such a coherent front state is 
constant in space and time in the  
frame that is comoving with the front velocity $v_F$ proved to be a good 
relaxation criterion to effectively determine whether such a state had been
obtained.

 \subsection{$+$Fronts}  \label{SEC:+front}

 \subsubsection{Structure and dynamics}  

\begin{figure}
 \includegraphics[clip=true,angle=0,width=14cm]{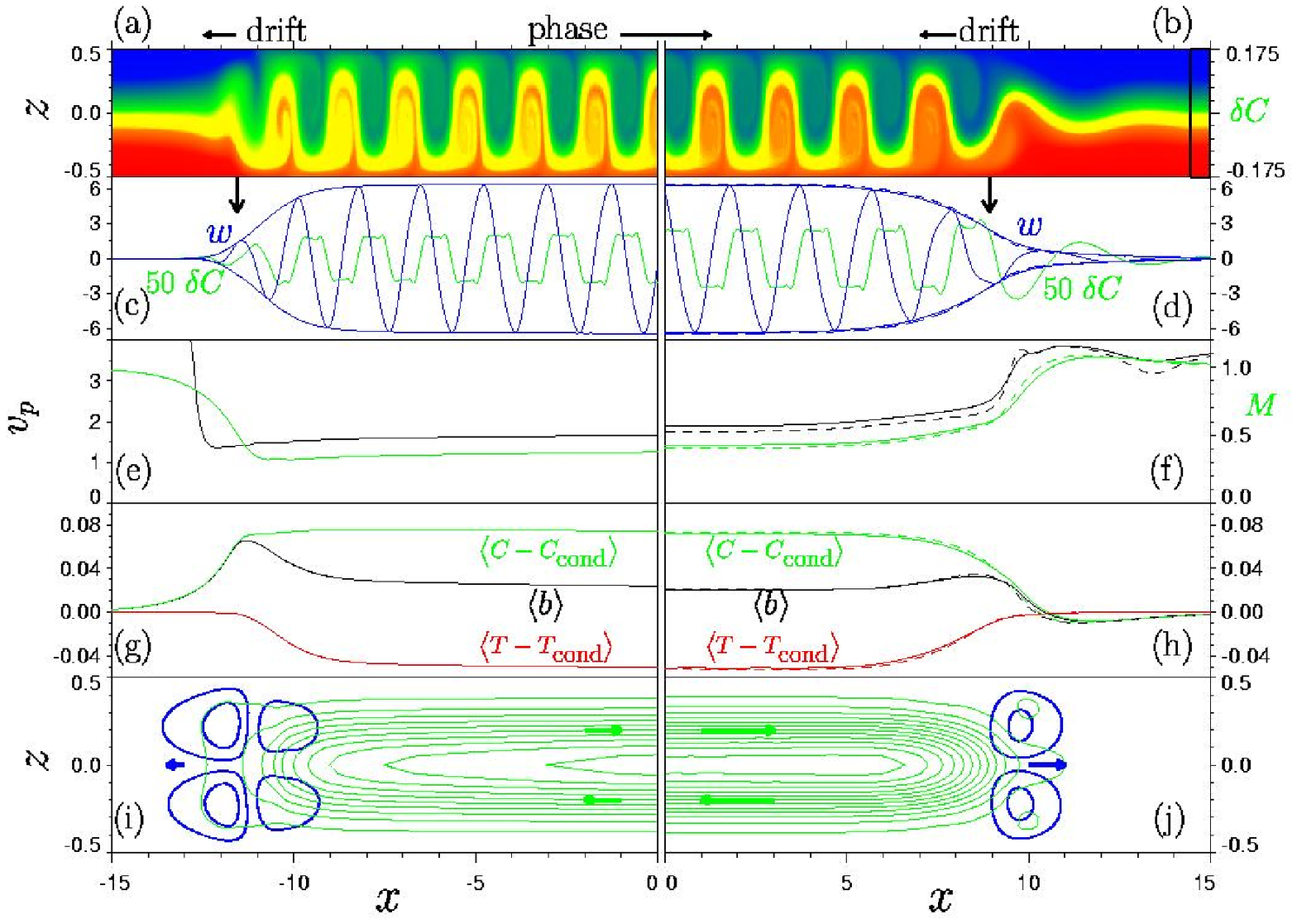}
 \caption{Typical structures of a coherent $+$front (left) and of a $-$front 
 (right) with nearly the same asymptotic wavelength $\lambda= 1.90$.
 Only the vicinity of the respective interfaces between
 convection and conduction is shown. 
 Both fronts propagate to the left ($v^+_F=-0.022, v_F^-=-0.067$), i.e., 
 opposite to the direction of the phase velocity $v_p$.
 (a, b) Color coded snapshot of concentration deviation $\delta C$ from its 
 global 
 mean in a vertical cross section of the layer. The color code is shown at the
 right end of (b).
 (c, d) Instantaneous lateral wave profile at midheight, $z=0$, of $\delta C$ 
 (green),  vertical velocity $w$ (blue), and its envelope.
 Arrows mark the positions where $w$ has grown up to $v_p$. 
 (e, f) Mixing number $M$ (green), Eq.~(\ref{Eq:M(x)}), and phase velocity 
 $v_p$ (black) of the nodes
 of $w$ in the comoving frames. The variation of $\lambda(x)=2\pi v_p(x)/\omega$
  is the
 same since the frequency $\omega$ is a {\it global} constant.
 (g, h) Time averaged deviations from the conductive state at $z=-0.25$ for 
 concentration
 (green), temperature (red), and their sum ($<b>$) measuring the convective 
 contribution to the buoyancy.
 (i, j) Streamlines of the averaged concentration current $<\vec{J}>$ (green) and
 velocity field $<\vec{u}>$ (blue). The latter results from $<b>$ and documents 
 roll shaped
 contributions of $<\vec{u}><\delta C>$ to $<\vec{J}>$ at the interfaces.
 Thick blue and green arrows indicate $<\vec{u}>$ and transport of positive
 $\delta C$ (alcohol surplus), respectively.
 Temporal averaging is always performed in the frame comoving with the 
 respective front velocity. Dashed lines show the decay
 part of the long LTW that coexists at the same parameters 
 ($r=1.3586, \psi=-0.35, L=0.01$) with the fronts. 
 Differences between the interfaces of the $+$front and the corresponding LTW 
 interface are not visible on the scale of the above plots.
 \label{Fig:frontpics}}
 \end{figure}

As soon as the growing convection rolls in a $+$front have become sufficiently 
nonlinear, i.e., when their lateral flow velocity $u$ has grown up to about 
their phase
velocity $v_p$ [e.g., close to the vertical arrow in Fig.~\ref{Fig:frontpics}(c)]
they start to alternatingly suck in positive ('blue') and negative ('red') 
$\delta C$ from the top and bottom concentration boundary layers, respectively. 
It is transported away into the well mixing convection bulk and replaced at the
interface location by neutral ('yellow/green') $\delta C$. Note that increasing
$u$ beyond $v_p$ causes the appearence of closed streamlines of the velocity
field in the frame comoving with the phase velocity of a traveling roll 
\cite{Linz88,Luecke98,JML04}. These closed streamlines regions
 are responsible for the characteristic roll
 structure of the $C$ field in Fig.~\ref{Fig:frontpics}(a): Positive (negative) 
 $\delta C$ is collected from the top (bottom) boundary
 layers and transported within the homogeneously mixed closed streamline 
 regions in phase direction while mean concentration, $\delta C \simeq 0$, is
 advected along the meandering "green-yellow stripe" in 
 Fig.~\ref{Fig:frontpics}(a) to the left \cite{Jung02}. The mean concentration 
 current   $<\vec{J}>$ resulting from this complicated concentration
 redistribution is shown in Fig.~\ref{Fig:frontpics}(i). All in all, mean
 concentration is accumulated (depleted) at the $+$ ($-$) front interface. 
 
The concentration redistribution reduces at the interface of the $+$front the
Soret-induced solutal 
stabilization that occurs to the left of it   
as a result of the large conductive vertical concentration gradient: at the
interface one can observe a minimal mixing number [Fig.~\ref{Fig:frontpics}(e)] 
and with it a buoyancy overshoot [Fig.~\ref{Fig:frontpics}(g)] which is 
sufficiently large to sustain local convection
growth there and cause even invasion of convection into the quiescent region whenever
$v_F^+<0$. With the fluid being well mixed there, i.e., with $M$ being small 
the local phase velocity is also small there -- in fact the minimum of $v_p(x)$
in Fig.~\ref{Fig:frontpics}(e) lies close to the one in $M$. 

Since the strongly stable quiescent fluid to the left of the $+$front 
prohibits a well developed advectively mixing 
front tail the reduction of $\delta C$ variations there is driven primarily by 
diffusion. 
The latter having a characteristic time scale given by  $L=0.01$ explains why
the front velocities are much smaller than the fast phase velocity.

When $r$ is increased $v_F^+$ tends to become (more) negative: 
convection to the right of the $+$interface can now, with increased heating, 
better invade the quiescent fluid to the left of it and thus 
$\partial_r v_F^+(r,\psi) < 0$. Similarly, when $\psi$ is increased, i.e.,
when the convection suppressing Soret effect is diminished the expansion of 
TW convection is favoured and thus $\partial_{\psi} v_F^+(r,\psi) < 0$.  

Moving along the $+$front in Fig.~\ref{Fig:frontpics} to the right 
from the interface towards the asymptotic TW state at large $x$ there develops 
an equilibrium  
between the $\delta C$ feed-in from the boundary layers at the plates and the
amount of advective mixing: The concentration contrast between two
neighboring rolls increases on the way towards the TW bulk. 
With it the phase speed $v_p(x)$, the wavelength 
$\lambda(x)=2\pi v_p(x)/\omega$, and the lateral concentration
current $<\vec{J}>$ grow monotonously up to their asymptotic TW values. This 
growth extends laterally over 
a wide interval which itself increases when the Soret coupling becomes stronger.
 
We found that the minimal wavelength in a $+$front state is located at the
interface and -- more remarkably -- that it is  about $\lambda_{min}\sim 1.4$ 
for {\it all $r$ and $\psi$} that we have simulated. We have no real 
quantitative explanation for this strong universal selection of the local
wavelength at the interface.
Intuitively the growing rolls are squeezed in the region with the negative
lateral gradient of $M$. The squeezing is relaxed when the rolls begin to 
absorb high concentration contrasts from the plate layers which increases 
$v_p$ again [arrow in Fig.~\ref{Fig:frontpics}(c)].

It is interesting to note that the mean concentration current $<\vec{J}>$ of TWs
becomes maximal close to the TW saddle node, i.e., where the asymptotic TW 
parts of our front states are located. 
Finally we mention that the front states do not sustain a measurable lateral 
meanflow; the quiescent fluid prohibits that.
On the other hand, extended TWs in laterally periodic systems show in general a
Reynolds stress-induced meanflow of the order $10^{-3}$ 
\cite{Linz88,Barten95I,Luecke98}. But it goes through zero just 
near the TW saddle node.

 \subsubsection{Bifurcation properties}  \label{SEC:bifprop}

\begin{figure}
 \includegraphics[clip=true,width=8.0cm,angle=0]{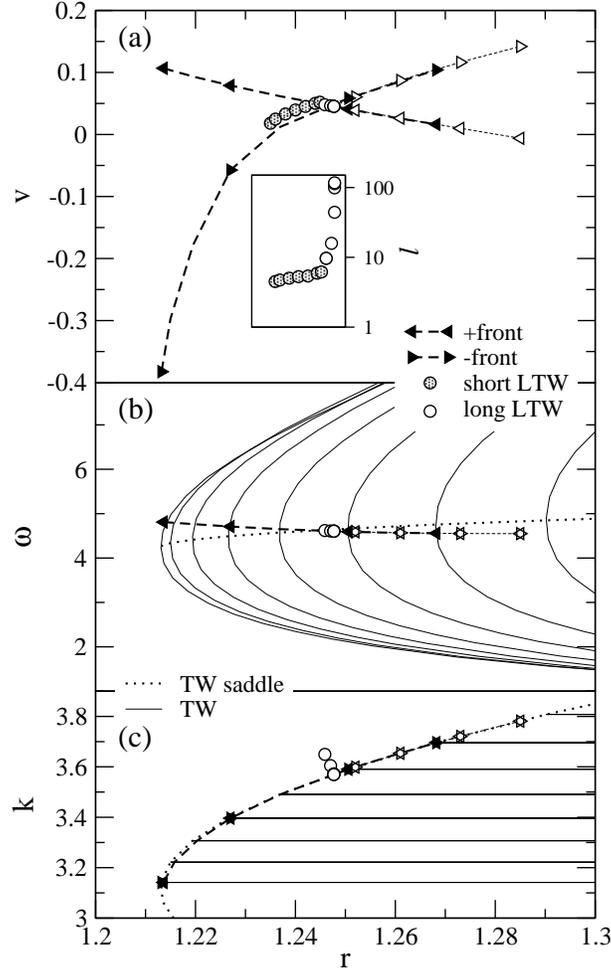}
 \caption{ Front and LTW bifurcation properties versus reduced Rayleigh number 
 $r=R/R^0_c$ for $\psi=-0.25$.
 Left and right pointing triangles with triangles denote $+$ and $-$fronts, 
 respectively. Open and shaded circles refer to long and  short LTWs, 
 respectively.
 (a) Front velocities of relaxed single-front states (thick dashed lines with 
 filled triangles), of
 expanding two-front states (thin dashed lines with open triangles), and drift 
 velocities of LTWs (circles).
 The inset shows the drastic increase of LTW length $l$ at
 $r^{LTW}_\infty=r^F_{eq}$ where the 
 front velocities of the $+$ and $-$ single-front states become equal.  
 (b) Frequencies of front states and of long LTWs in comparison with the rest 
 frame frequencies of laterally periodic TWs. The saddle-node
 vicinities of the latter are shown by full lines for several wave numbers 
 $k=2\pi/\lambda$. TW states with frequencies above the dotted line of 
 saddle-node TWs are unstable.   
 (c) Wave numbers selected by front states in the bulk part far away from the
 interface (triangles) and in the central part of long LTWs (open circles). 
 Horizontal lines indicate the laterally periodic TWs that are shown in (b) by
 full lines. The continuum of these TW states is bounded in the $r-k$ plane by 
 the dotted line of TW saddle nodes, 
 $k_s^{TW}(r)$. Here we show only the large-$k$ branch of it (cf. the dotted
 line marked $r_s^{TW}$ in Fig.~\ref{Fig:3DPsi40om-k-r} for another 
 perspective).
 \label{Fig:Psi25v+om+k-r}}
 \end{figure}

In Figs.~\ref{Fig:Psi25v+om+k-r} - \ref{Fig:3DPsi40om-k-r} we show the
bifurcation properties of fronts in comparison with LTWs and laterally periodic
TW states. We use front velocities and frequencies being temporally and 
spatially constant as order parameters to characterize all of the aforementioned 
oscillatory states. In addition we also consider the local wave numbers of 
front states and of LTWs in the bulk spatial regions where $\lambda(x)$ has 
reached a plateau, i.e., sufficiently away from any interface to conduction.

Figs.~\ref{Fig:Psi25v+om+k-r} and \ref{Fig:3Psisv+l+w-r} show that the front
velocities of $+$ and $-$fronts vary quite differently as a function of $r$.
The former decrease linearly with growing $r$ and the latter increase, albeit
not linearly. Thus, there is a crossing at $r^F_{eq}$ where $v_F^+$ becomes 
equal to $v_F^-$, so that both fronts move with the same velocity. At this
Rayleigh number the length $l$ of the LTWs diverges, i.e., 
$r^{LTW}_\infty = r^F_{eq}$. There, and strictly speaking only there, this 
limiting LTW can be seen as a state consisting of two fronts. 

The frequency and bulk wave number selected by a $+$front and of a very long 
LTW are close to those of the respective, laterally extended saddle-node 
TW (Figs.~\ref{Fig:Psi25v+om+k-r}, \ref{Fig:4Psisom+r-k}, \ref{Fig:3DPsi40om-k-r}).   
A somewhat hand-waving explanation for the selection of the saddle-node
frequency is as follows: With ({\it i}) convection growing out of conduction in a $+$
front, with ({\it ii}) small-amplitude extended TW perturbations of the latter 
oscillating according to a purely linear balance with the large Hopf frequency, and 
with ({\it iii}) the tendency to decrease 
$\omega$ with
growing convection amplitude the saddle-node frequency is the first, i.e., the
largest possible eigenfrequency of the full nonlinear front problem to allow 
for a stable TW region away from the interface.

A stable front state that has a TW bulk part extending laterally to infinity 
with frequency $\omega$ and wave number $k$ cannot be realized at $r$-values 
that lie below the saddle-node curve of laterally
extended TWs, cf. the curve marked $r_s^{TW}$ in the $k-r$ plane of 
Fig.~\ref{Fig:3DPsi40om-k-r}. Thus, the lowest Rayleigh number $r^F_{min}$ for
the existence of fronts is $r_{min}^{TW} = r_s^{TW}(k \simeq \pi)$, i.e., the 
location of the
tip of the nose-shaped TW bifurcation surface like the grey surface in 
Fig.~\ref{Fig:3DPsi40om-k-r}. Ahead of this nose one
cannot realize front states because at such locations there are no 
TWs to which the interface from conduction could connect. 

The TW bulk parts of our $+$fronts are practically saddle-node TWs that have  
bulk wave numbers on the saddle-node curve $r_s^{TW}(k)$.
Furthermore, it is interesting to note that they are on the {\em large}-$k$ 
branch of 
$r_s^{TW}(k)$ --- the big plusses in Fig.~\ref{Fig:3DPsi40om-k-r} marking the
bulk values of the front states lie all above $k \simeq \pi$. In fact, in all 
our simulations we did not find front states with bulk wave numbers smaller than 
$\pi$. This value marks for all $\psi$ that we investigated the tip of the 
nose-shaped TW bifurcation surface like the grey surface in 
Fig.~\ref{Fig:3DPsi40om-k-r}.

In contrast to fronts, however, LTWs of 
{\em finite} length $l$ can coexist bistably
together with the conductive state at $r$-values well below $r_s^{TW}(k)$: They 
can sustain over a finite lateral length convection with frequencies and bulk
wave numbers (big bullets in Fig.~\ref{Fig:3DPsi40om-k-r}) ''ahead'' of the 
grey TW surface for reasons that are explained in Ref.~\cite{Jung02}. 
This also shows that fronts and LTWs are quite different states. In the 
limit $l \to \infty$ the LTW states merge at $r^{LTW}_\infty=r^F_{eq}$ with a 
TW whose wave number and frequency is close to the TW saddle-node as shown in 
Figs.~\ref{Fig:Psi25v+om+k-r}, \ref{Fig:4Psisom+r-k}, \ref{Fig:3DPsi40om-k-r}.
Therefore, $\omega(r^{LTW}_\infty)$ increases when the Soret coupling becomes 
more negative but $k(r^{LTW}_\infty)$ decreases. For $\psi \lesssim -0.4$ it 
moves towards the tip of the TW nose at $k\simeq \pi$.

 \subsubsection{Front selection and stability}  

Simulations of $+$fronts that were done at fixed control parameters 
$r, \psi$ with different initial conditions, e.g., different 
wave numbers of an initial TW part produced in general a uniquely 
selected final front state with the same frequency and the same 
asymptotic bulk TW part.
During the formation process initial wave structures with the 'wrong' wave
patterns propagated out of 
the system in the direction of the phase velocity and 
were substituted by convection that was selected by the front. That also 
explains why our TW boundary condition $f(L)=f(L-\lambda)$ at the 'downstream'
end has no measurable influence on the $+$front state even 
when $\lambda$ differs from the front-selected value.

The substitution dynamics is documented in Fig.~\ref{Fig:frontrelaxation}. 
There a TW bulk part was prepared initially at $x>8$ with a wavelength of 
$\lambda=1.85$ and phase velocity $v_p=1.032$. The spatial region to the right
of the interface to conduction is then invaded by the front-selected TW pattern
that has a smaller bulk wavelength of $\lambda=1.80$ and that propagates with a 
faster phase velocity of $v_p=1.258$. The wave number is increased via several 
phase annihilating defects.

All our $+$interfaces selected bulk TW wave numbers close to the large-$k$ 
branch of the TW saddle-node curve; cf. Figs.~\ref{Fig:4Psisom+r-k}(b) and 
\ref{Fig:3DPsi40om-k-r} for $r^{TW}_s(k)$ and Fig.~\ref{Fig:Psi25v+om+k-r}(c)
for $k^{TW}_s(r)$. Thus, these wave numbers are too large 
to be Eckhaus stable \cite{Baxter92,Kolodner92,Bu99,MeAlBa04}.
However, these fully developed TWs were only convectively unstable 
\cite{Bu99}: Perturbations could grow but while doing so they
were advected sufficiently fast downstream in the direction of the TW phase 
propagation so that they could not influence the upstream part of the $+$front 
state in a persistent way. In systems with sufficiently long
downstream section of the front state the growing fluctuations have sufficient 
time --- or are sufficiently fast growing, respectively --- to reach a critical
 amplitude at which
two neighboring rolls are annihilated \cite{Baxter92,Kolodner92} as, e.g., for
the parameters of Fig.~\ref{Fig:irregularfront}.

Because noise cannot be prevented in general one observes then 
such phase defects as in Fig.~\ref{Fig:irregularfront} at irregular points in 
time and space beyond a certain downstream growth length that is related to 
size of the noise and the growth rate. The associated roll-annihilation events 
can lead to an effectively reduced mean wave number in the very far downstream
region of the convection bulk. Thus, the coherent part of the $+$front 
close to the interface to conduction is followed by a second 
incoherent, chaotic phase front consisting of the erratically occurring
phase defects. This phase front connects the smooth primary Eckhaus unstable 
section to a smooth
Eckhaus stable TW with smaller wave number that is realized at larger downstream
distances.
For parameters for which the growth rate of perturbations of the primary
front-selected TW is lower than the one of Fig.~\ref{Fig:irregularfront}
one does not observe in short systems the erratically occurring phase defects
 --- and even less so the Eckhaus stable final downstream TW state. Indeed, 
that was the situation for most of our front states. 

We finally mention that we could also generate front states with frequencies
larger than those of the laterally extended saddle node TWs [dotted lines in 
Fig.~\ref{Fig:Psi25v+om+k-r}(b) and Fig.~\ref{Fig:4Psisom+r-k}(a)], i.e., with  
frequencies that lie above the respective dotted line in the
respective 3D plot similar to the one of Fig.~\ref{Fig:3DPsi40om-k-r}. However,
we suppose that in sufficient long systems and after long enough times
these unstable TW realizations develop a $-$front in the downstream bulk 
possibly induced by roll annihilating defects \cite{Kolodner92}.

 \subsection{$-$Fronts}  

The right half of Fig.~\ref{Fig:frontpics}(b) shows a typical $-$front. The
mean lateral concentration current in the TW bulk part to the left of the 
$-$interface to conduction shuffles positive (negative) $\delta C$ in the upper
(lower) part of the layer towards the $-$interface. Thus, a large vertical
concentration gradient is maintained slightly ahead of it that strongly 
stabilizes the conduction regime to the right of the $-$interface:
there the mixing number $M$ is even larger than 1. In this way the TW
oscillations are damped and the conduction regime is shielded against a rapid
invasion of convection. 

The increase of $M(x)$ upon approaching the interface from the convection side 
causes --- and is related to --- a similar increase of $v_p(x)$ and $\lambda(x)$.
The rolls disperse with growing phase velocity $v_p(x)$ over a short lateral 
distance at the interface. The decreasing convection amplitude lowers the mean 
concentration current and causes $M(x)$ to grow further. This in turn enhances 
$v_p(x)$ and $\lambda(x)$ leading to 
smaller convection amplitude and so on. It is therefore the strongly nonlinear 
lateral concentration current $<\vec{J}>$ which is responsible for the 
rapid self amplified decay process of convection at the $-$interface.

With increasing $r$ the front velocity $v_F^-$ changes sign, becomes positive
and continues to grow [Fig.~\ref{Fig:Psi25v+om+k-r}(a) and 
Fig.~\ref{Fig:3Psisv+l+w-r}(a)] because the quiescent state becomes
less stable when increasing $r$.
The slope of $v_F^-(r)$, i.e., the increase of the front velocity is 
considerably steeper for negative $v_F^-$ than for positive ones:
The strongly stabilizing solutal stratification ahead of the interface hinders 
convection to intrude
into the quiescent fluid region but favours the latter to replace the TW 
part.

The 'upstream' lateral distance over which the $-$interface to conduction 
influences the TW to the left of it is definitely smaller than the 'downstream'
influence length of the $+$interface on the convective bulk. In the former case
one cannot observe 
a difference to an extended TW state at an 'upstream' distance of, say, 10-15 
rolls while in the latter case the 'downstream' convection properties
approach the asymptotic bulk TW behavior only over a significantly longer 
distance. So, in particular the phase dilatation at a $-$interface does not  
propagate 'upstream' into the TW bulk against the fast phase flow.

This also explains why in the formation process of a $-$front TW 
properties that were initially present in a developed form are conserved. In 
fact, we could 
produce coherent $-$fronts for a fixed $r$ with different wave numbers of the 
bulk TW part out of a whole band near the saddle node wave numbers 
$k_s^{TW}(r)$. Only for higher $r$ and initial wave numbers away from 
$k_s^{TW}$ we observed long-time transient incoherent front behavior.
Here this transition to incoherence may correspond to a transition from 
a convectively to an absolutely unstable regime concerning the propagation of 
phase dilatations in 'upstream' direction.

We should like to stress again that in contrast to $-$fronts which depend on
the preparation process the asymptotic 
'downstream' TW part of a $+$front is uniquely selected as discussed in the
previous section. Thus, for a particular $r$ we have found only a single 
coherent $+$front.
 
For definiteness and for the sake of comparison with $+$fronts we show in the 
Figs. of this paper the properties of $-$front states that have a bulk TW part
which itself was selected by a $+$front.
This, however, has a slight numerical drawback stemming from the convective
Eckhaus instability of this TW part: ever present phase noise (in particular 
at the boundary of the 'upstream' TW region) is enhanced on the 'downstream' 
way towards the $-$interface. We think that this effect is 
responsible for fluctuations in our frequency measurements of $-$fronts. These 
data are therefore not shown in Figs.~\ref{Fig:Psi25v+om+k-r}(b) and 
Fig.~\ref{Fig:4Psisom+r-k}. However, in our simulations the 'upstream' part of 
the $-$fronts were too short to allow for the full development of phase slip 
defects.

 \subsection{Transient two-front structures}  

\begin{figure}
 \includegraphics[clip=true, angle=0,width=8.5cm]{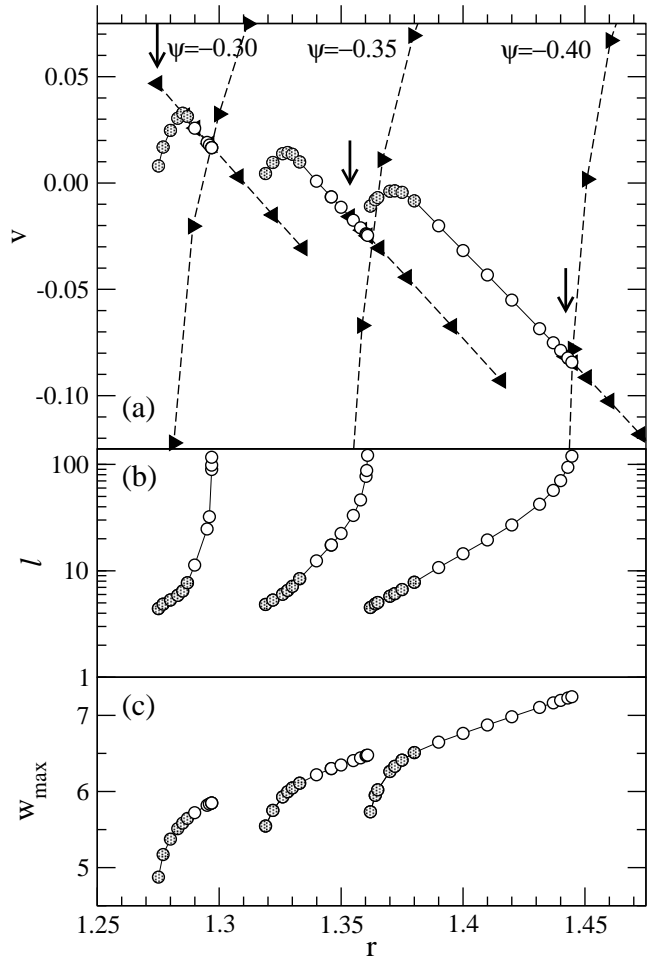}
 \caption{ Front states and LTWs for different $\psi$.
 (a) Front velocities, $v_F$, of $+$fronts (left pointing triangles), of  $-$ 
 fronts (right pointing triangles), and drift velocities, $v_d$, of LTWs
 (circles) versus $r$. Note that short LTWs (shaded circles) and long LTWs 
 (open circles) show different $v_d(r)$-behavior.
 The latter varying linearly with $r$ is nearly indistinguishable from $v^+_F(r)$. 
 Arrows mark the low-$r$ existence boundary $r_{min}^{TW}$ of laterally 
 periodic TWs
 and with it of front states. LTWs exist below this threshold \cite{Jung02} with
 drift velocities that show the above mentioned linear variation
 with $r$ as long as the LTWs are long enough. To identify an LTW as a long one
 we required a clearly visible plateau in the spatial properties. 
 (b) Length $l$ of the LTWs of (a) measured as the distance between 
 the half maximum values of the envelope of the vertical velocity field $w$ 
 [cf., blue line in Fig.~\ref{Fig:frontpics}(c,d)].
 (c) Maximal vertical flow velocities $w_{max}$ of LTWs.
 \label{Fig:3Psisv+l+w-r}}
 \end{figure}

\begin{figure}
 \includegraphics[clip=true, angle=0,width=8.5cm]{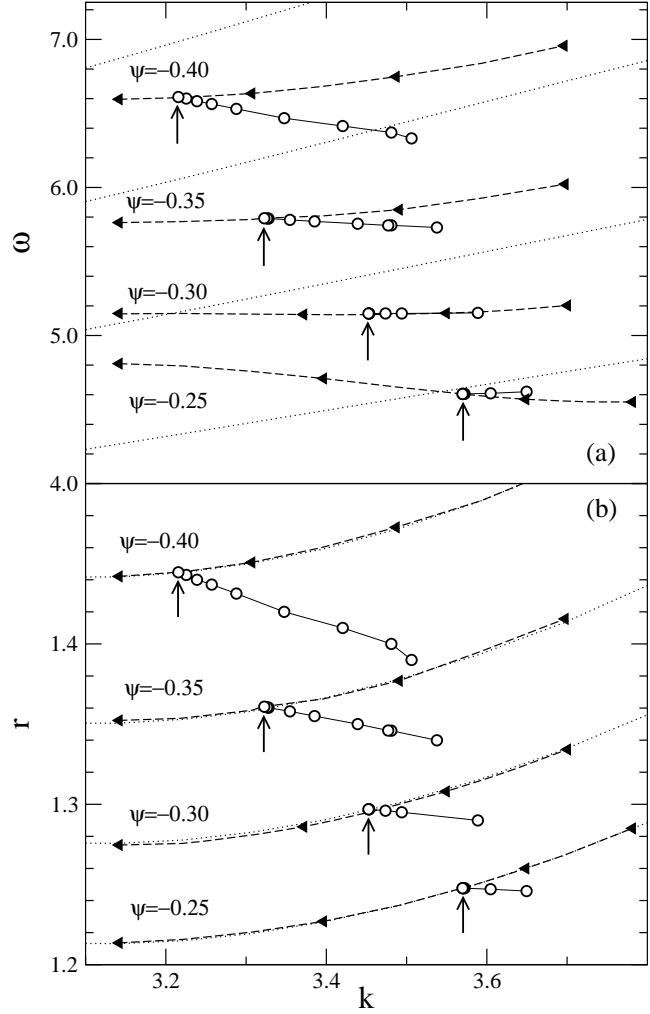}
 \caption{ Wave properties of $+$fronts (dashed lines with triangles), 
 long LTWs (circles), and laterally extended saddle-node TWs (dotted 
 lines) for different $\psi$. The wave numbers of the two former refer to
 plateau values in the bulk part away from the interface. 
 (a) Frequency $\omega$ (for TWs in the rest frame and for LTWs and fronts in
 the comoving frame) versus wave number $k$.
 (b) The same convection structures in the $k-r$ plane.
 The wave numbers and frequencies of LTWs with
 $l \to \infty$ (arrows) coincide at $r^{LTW}_\infty=r^F_{eq}$ with those of 
 the respective $+$  fronts.
 \label{Fig:4Psisom+r-k} }
 \end{figure}

\begin{figure}
 \includegraphics[clip=true, angle=-90,width=8.5cm]{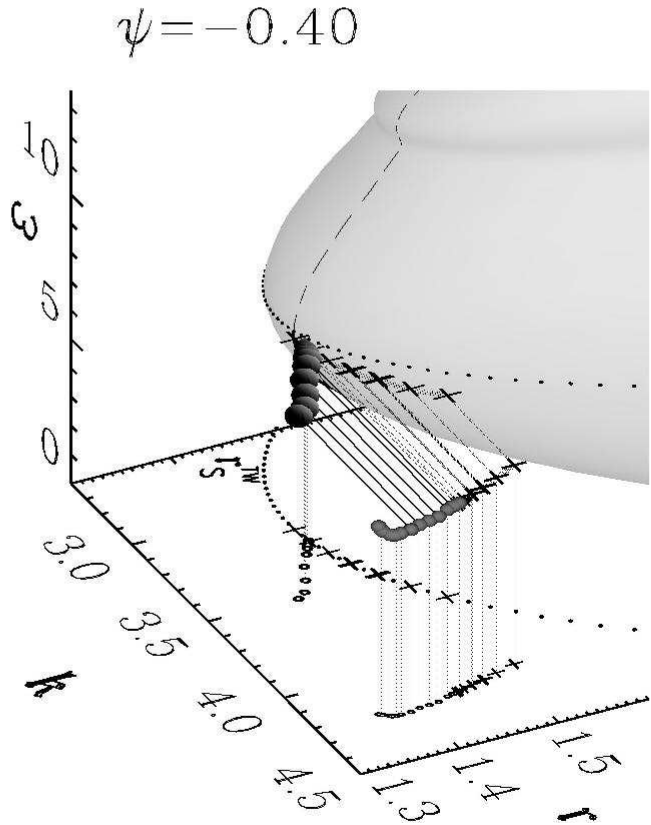}
 \caption{ Laterally periodic TWs, $+$fronts, and LTWs
 in the three dimensional $k-r-\omega$ parameter space.
 Grey, nose-shaped surface \cite{Jung02} denotes TWs. They are unstable when
 $\omega$ is above the dotted line of saddle nodes. Its projection onto the
 $k-r$ is marked by $r^{TW}_s(k)$. A particular bifurcation branch for a given 
 $k$ (e.g., the long-dashed line for $k\sim \pi$) starts backwards with a large
 Hopf frequency (not shown) and becomes stable by a saddle-node bifurcation at 
 the dotted line.
 The big plus signs on the TW surface mark asymptotic $+$fronts. The small
 plusses at large $k$ denote the 
 highest wave numbers occurring at the $+$front interface. Long LTWs are 
 represented with their plateau values by big bullets. 
 They coexist with fronts (big plusses) in a very narrow $r$ interval at 
 $r^{LTW}_\infty$ close to the tip of the TW nose. The small bullets at large $k$ 
 denote the largest local wave numbers occurring at the interface of convection
 growth. Each horizontal line indicates at fixed $r$ and $\omega$ the spatial 
 variation 
 of the local wave number $k$ within a $+$front or a long LTW from the growth 
 interface to the asymptotic plateau value. The $k$ variation of LTWs from 
 plateau to the decay interface into conduction is not shown.
 \label{Fig:3DPsi40om-k-r}}
 \end{figure}

Consider a set-up where a $+$front generates a very long TW part that develops 
back into the quiescent fluid via a coherent $-$front. If the convective 
part is laterally sufficiently long then this structure appears as a two-front 
structure with the TW part being spatially confined between a $+$ and a 
$-$interface to conduction. This structure will in general either expand or
shrink laterally. Only when the two front velocities $v_F^+$ and $v_F^-$ are 
equal, i.e., at the
crossing points $r^F_{eq}$ of the curves in Figs.~\ref{Fig:Psi25v+om+k-r}(a) and
\ref{Fig:3Psisv+l+w-r}(a) one can have a stationary state, namely, 
a LTW with diverging length $l$.

We have simulated such structures at $\psi=-0.25$ and $-0.30$ for Rayleigh 
numbers for which $v_F^+<v_F^-$, i.e., to the right of the crossing in 
Figs.~\ref{Fig:Psi25v+om+k-r}(a) and \ref{Fig:3Psisv+l+w-r}(a) so that these 
two-front structures expand. Their properties are practically identical 
to those of the respective single-front states. The two-front structures
have a technical advantage over the simulation of  single fronts:
we could use a periodic boundary condition that was located in the quiescent
region of the former. This avoids the noise that is induced at the upstream TW 
boundary of a single $-$front state. With this noise source being absent in our 
two-front structures the frequency fluctuations at the $-$
interfaces of single $-$front states did not occur. 

When we reduced $r$ then the velocities of the two
fronts approached each other along the lines in 
Figs.~\ref{Fig:Psi25v+om+k-r}(a) and \ref{Fig:3Psisv+l+w-r}(a) that were
obtained from velocity measurements of single-front states. In that way we could
reproduce the unique crossing point at $r^F_{eq}$ where $v_F^+=v_F^-$ and
where LTWs with diverging $l$ exist with a drift velocity $v_d$ given by the
crossing velocity.

In addition to expanding two-front structures we simulated also shrinking ones
for a particular parameter combination ($\psi=-0.35, r=1.3586$, for which 
$v_F^+=-0.022 > v_F^-=-0.067$) that is located in Fig.~\ref{Fig:3Psisv+l+w-r}(a)
at $r<r^F_{eq}$. As an example consider the 
large two-front structure as in Fig.~\ref{Fig:LTWapproach}. 
The $+$front selects in the bulk of this initial two-front structure a saddle 
 node TW with wavelength $\lambda_{plateau}\sim 1.905$. In the course of time 
 the velocity  $v_F^-$ of the $-$front 
approaches that of the $+$front and a stationary LTW forms with length 
$l \simeq 47$ that drifts with the velocity $v_d=v_F^+=-0.022$ and oscillates
with the frequency of the $+$front. These values of the $+$front remain
practically unchanged in the whole process. 

\section{Localized traveling wave states} \label{SEC:LTW}

We produced LTW states with very large length $l$ immediately below the 
Rayleigh number $r^{LTW}_\infty = r^F_{eq}$. There, $l$ diverges thus marking 
the upper existence boundary of LTW states. And there, the 
velocities $v_F^+$ and $v_F^-$ of the $+$ and $-$front states, respectively, 
become equal. See Figs.~\ref{Fig:Psi25v+om+k-r} and \ref{Fig:3Psisv+l+w-r} for
the corresponding results in the range of $-0.4 \leq \psi \leq -0.25$ that we
have investigated in this paper. Upon decreasing $r$ below the threshold 
$r^{LTW}_\infty(\psi)$ one finds uniquely selected 
LTW states. Depending on parameters they can coexist stably with front states,
extended TW states, and the quiescent basic state.

For completeness we include here in Fig.~\ref{Fig:phasediagram} a phase 
diagram of the $\psi -r$ plane where
all the LTWs that we have numerically obtained in the range 
$-0.65 \leq \psi \leq - 0.08$ are shown by vertical bars together
with the saddle-node location $r_s^{TW} (k = \pi)$ of extended TWs (full line)
and the oscillatory Hopf bifurcation threshold $r_{osc}(k = \pi)$ for TWs 
(dashed line). But in this work we focus on the range 
$-0.4 \leq \psi \leq -0.25$.

We should like to emphasize again that we found LTWs for 
$-0.4 \leq \psi \leq -0.25$ only in the parameter regime below 
$r^{LTW}_\infty(\psi)$, i.e.,  
to the left of the crossing points in Figs.~\ref{Fig:Psi25v+om+k-r}(a) and 
\ref{Fig:3Psisv+l+w-r}(a) where the velocities of independent single fronts 
become equal. Thus, LTWs exist at these $\psi$ only for parameters for which 
$v_F^+ > v_F^-$, 
i.e., for which independent fronts would approach each other: eventually any 
convective region between them would shrink to zero and the quiescent conductive
state would result if this interface motion would continue without change. 
However,
the stabilization effects that allow in such a situation a uniquely selected 
stable and robust LTW are easily understood with the help of the investigations
in the following subsections. On the other hand, for $r>r^F_{eq}$ the front
velocities are such that a two-front structure expands.

\subsection{Transient dynamics towards the selected LTW}

\begin{figure}
 \includegraphics[clip=true, width=15cm]{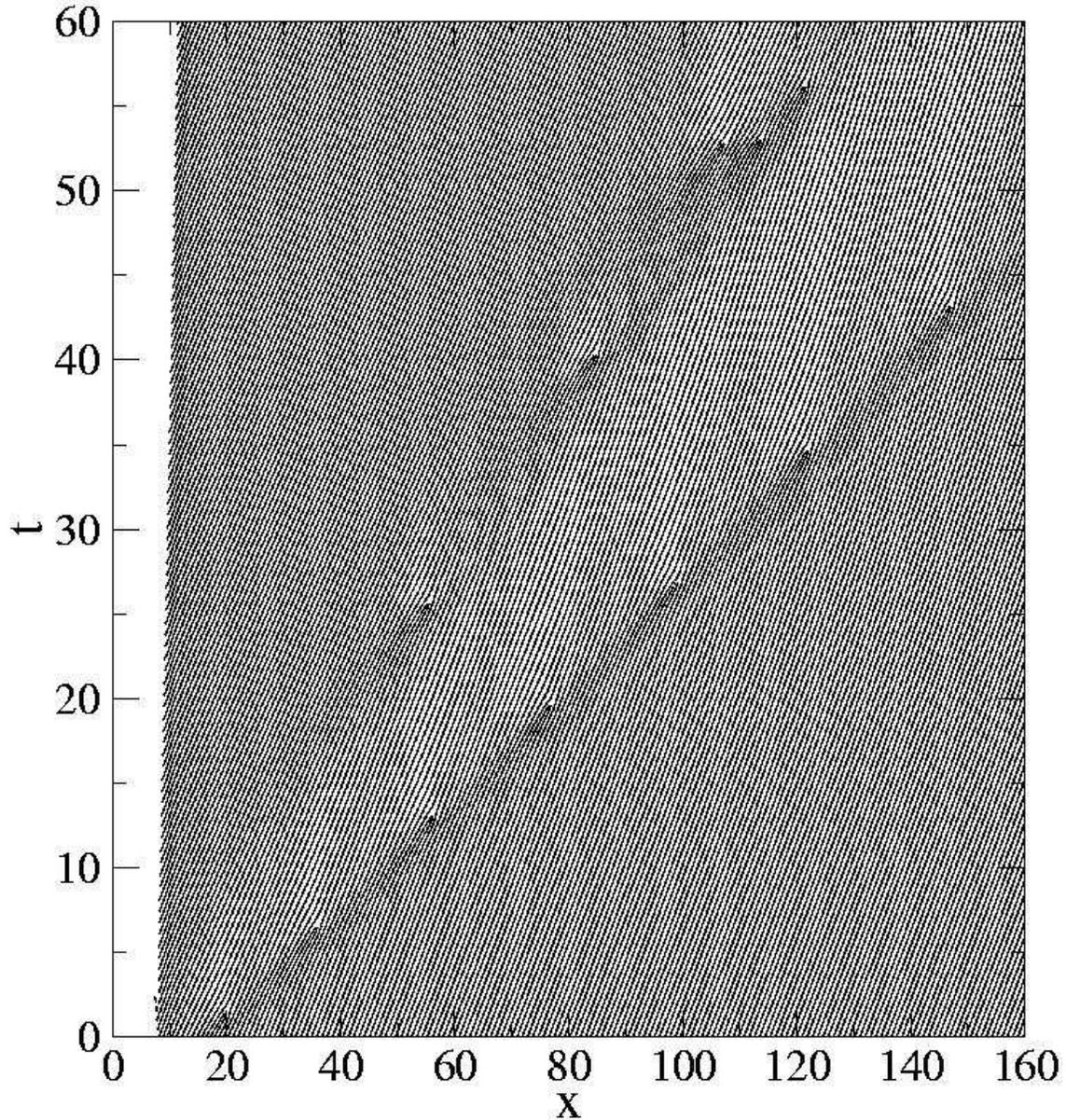}
 \caption{ Typical spatio-temporal evolution of a $+$front. Shown are the 
 extrema positions of the vertical velocity field $w$. 
 The initial condition at time $t=-5$ (not visible) consisted of an extended TW 
 for $x>8$ with wavelength $\lambda=1.85$ and phase velocity $v_p=1.032$ 
 and quiescent fluid for $x<8$.
 Boundary conditions are conduction at $x=0$ and $f(x=160)= f(x=160-1.85)$ 
 that imposes at $x=160$ a wavelength of $\lambda=1.85$.
 First, a pulse that causes a quite regular sequence of roll-pair annihilation 
 events (lower line of defects) propagates to the right with velocity greater 
 than 
 $v_p$. The intermediate wave pattern resulting from this primary sequence of
 defects is then transformed via further, somewhat erratically occurring
 phase defects into the fast asymptotic TW with $\lambda=1.80,v_p=1.258$ that is
 favoured by the $+$front. The boundary condition 
 at $x=160$ that imposes a 'wrong' wavelength there does not influence the 
 bulk behavior which is selected by the front.
 Parameters are $r=1.237, \psi=-0.25$. 
 \label{Fig:frontrelaxation}}
 \end{figure}

\begin{figure}
 \includegraphics[clip=true,width=15cm]{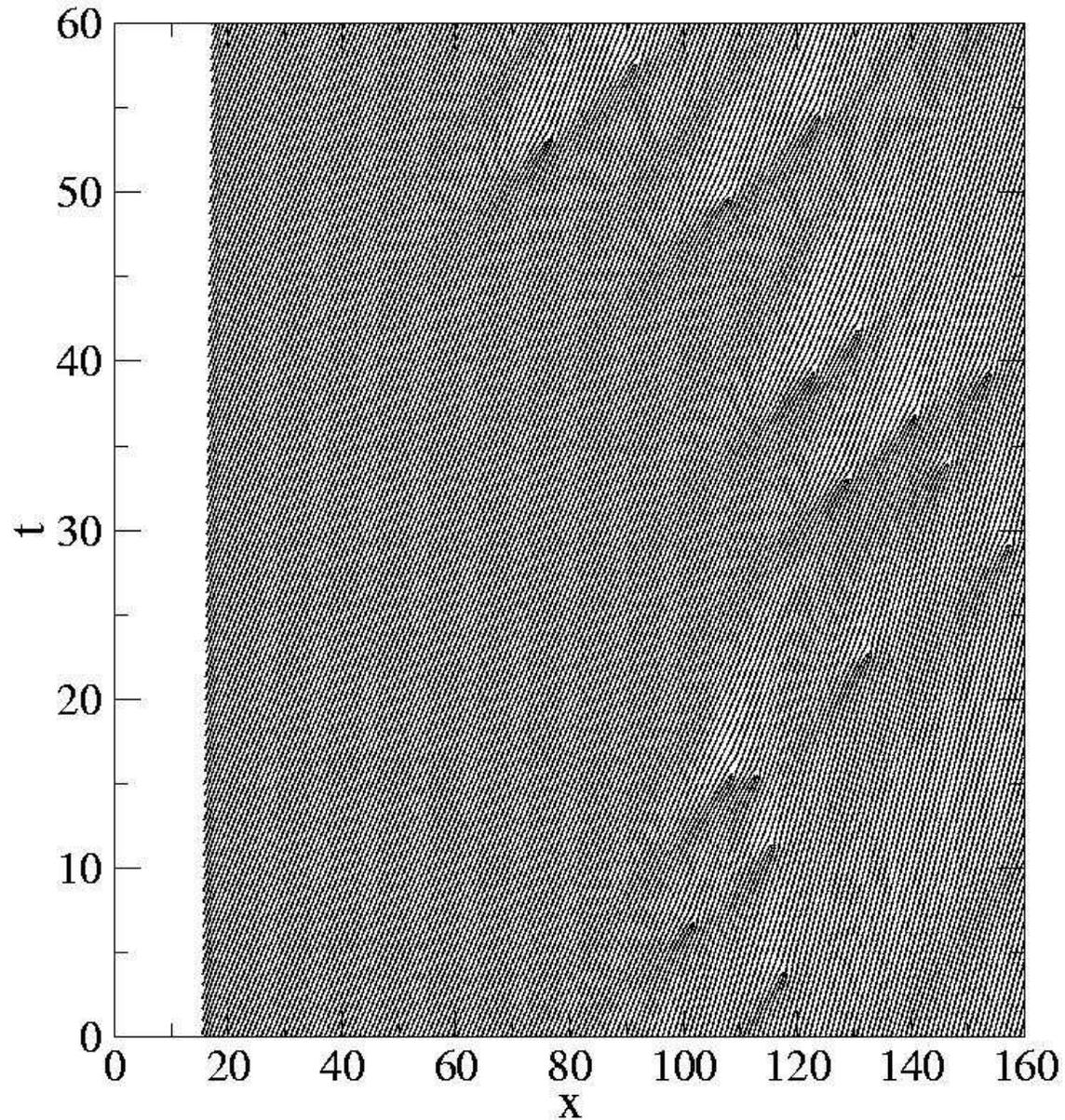}
 \caption{ Spatio-temporal dynamics of a front-selected TW pattern that is 
 convectively Eckhaus unstable. Shown are the  extrema positions of the 
 vertical velocity field $w$. The front selects a bulk TW 
 with wavelength $\lambda \sim 1.72$ that is strongly Eckhaus unstable:
 While beeing advected 'downstream' phase deformations (that are caused, e.g.,
 by computer 'noise') grow sufficiently fast to reach a critical amplitude 
 within the system length. Then two neighboring rolls are annihilated. That 
 increases the wavelength and decreases the phase velocity towards Eckhaus 
 stable values. Parameters are $r=1.26, \psi=-0.25$.
 \label{Fig:irregularfront}}
 \end{figure}

\begin{figure}
 \includegraphics[clip=true, angle=0, width=8.5cm]{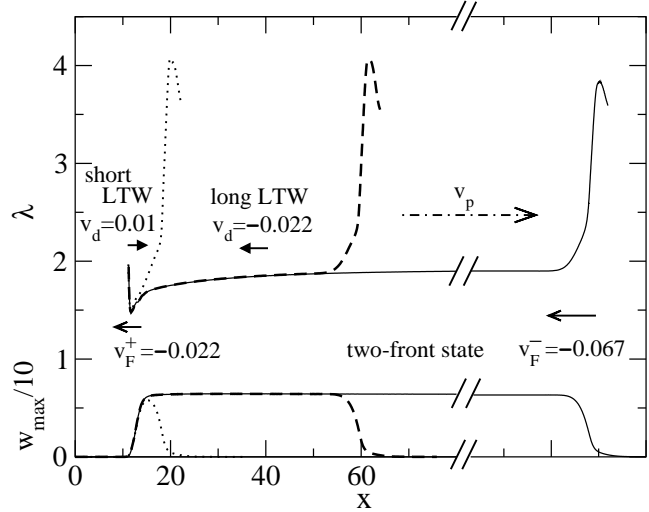}
 \caption{ Evolution of the lateral profiles of the wavelength (top) and of 
 the vertical flow amplitude $w_{max}/10$ (bottom) after 
 starting with a very long two-front structure at $r=1.3586, \psi=-0.35$. 
 The $+$front selects in the bulk of the initial two-front structure a 
 saddle-node TW with wavelength $\lambda_{plateau}\sim 1.905$. With  
 $v^-_F<v^+_F<0$
 the $-$front approaches the $+$front and doing so the velocity of the former
 goes monotonously towards $v_F^+$. This transient process ends in a long LTW 
 (dashed line) of constant length $l \simeq 47$ with a plateau wavelength of $1.873$.
 Its drift and frequency is effectively the same as the respective values of 
 the $+$front which remain unchanged all the time.
 For comparison the profiles of a short LTW at the same $\psi$ but smaller  
 $r=1.3220$ are shown with dotted lines.
              \label{Fig:LTWapproach}}
 \end{figure}

\begin{figure}
 \includegraphics[clip=true, angle=0, width=8.5cm]{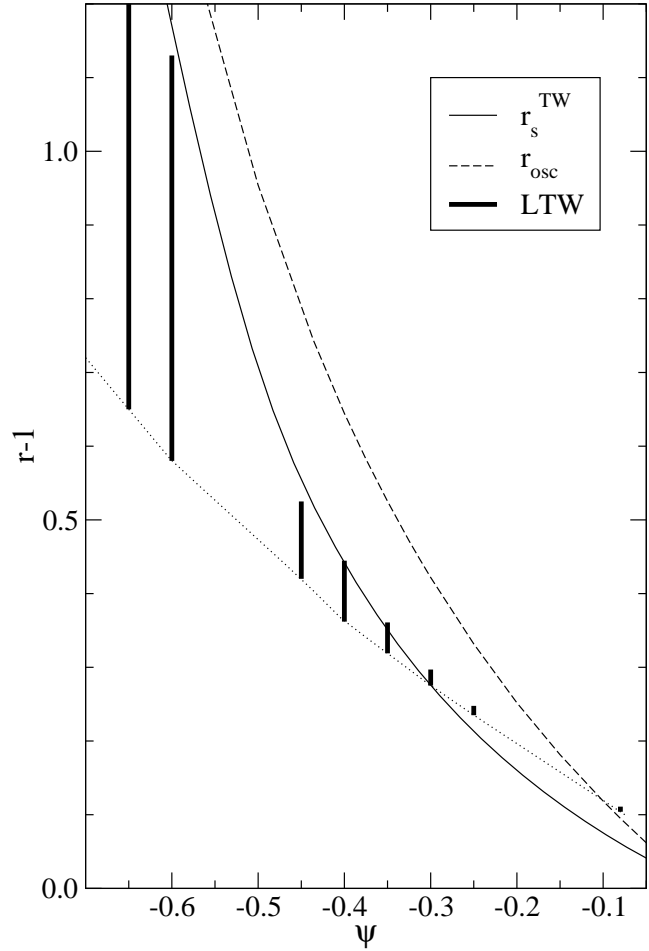}
 \caption{Phase diagram in the $\psi -r$ plane. The vertical bars indicate 
the range of stable existence of those LTWs that we
have numerically simulated. Full and dashed lines refer to the saddle node
location $r_s^{TW}$ of extended TWs and to their 
oscillatory Hopf bifurcation threshold $r_{osc}$, respectively; both for a 
wave number $k = \pi$ . For $\psi \leq -0.25$ the upper existence boundary of
LTWs was determined by the requirement that $l$ remained below about 120 in 
our numerical set-up. The dotted line guides the eye along the lower 
band limit $r^{LTW}_{min}$ of LTWs. Parameters are $L = 0.01, \sigma = 10$.
 \label{Fig:phasediagram}}
 \end{figure}

A typical transient dynamics towards the uniquely selected LTW is shown in 
Fig.~\ref{Fig:frontrelaxation} for $\psi=-0.35$. Here the initial condition was a 
very broad two-front structure that was prepared at 
$r=1.3586$ where it shrinks with $v_F^+ > v_F^-$ [Fig.~\ref{Fig:3Psisv+l+w-r}(a)]. 
In fact, the $-$front moves to the left with a speed that is about three times 
higher than that of the $+$front. 

In the following shrinking process 
where the $-$front closes up to the $+$front the latter does not change its
velocity at all and the former keeps its velocity as long as 
the bulk TW part between the two fronts is effectively asymptotic, i.e., without 
lateral variation. This behavior reflects the fact that in such broad two-front
structures there is practically no interaction between the fronts when
their distance is so large that an asymptotic TW part is realized between the 
interfaces to conduction. 
However, the situation changes when the convective region between the interfaces 
becomes less extended since
it requires a finite 'downstream' growth length behind a $+$interface over 
which the 
convection properties still vary with small gradients before the asymptotic TW 
is reached. The slow lateral variation is best seen in the mixing
number $M(x)$ in Fig.~\ref{Fig:frontpics}(e) reflecting the slow variation of 
the concentration distribution and in the related convective contribution 
$\langle b \rangle$ [Fig.~\ref{Fig:frontpics}(g)] to the local buoyancy.

When the front separation comes to the order of this length
one cannot speak any more of a two-front structure with two independent 
fronts: For the parameters of Fig.~\ref{Fig:frontrelaxation} the velocity 
of the formerly independent $-$front changes continuously from $v_F^-=-0.067$ 
to $-0.022$, i.e, to the velocity of the preceeding $+$front and a coherent and
robust LTW 
forms which moves with a drift velocity that is determined by the $+$front,
$v_d = v_F^+=-0.022$. During this slowing-down process of the $-$interface 
its structure changes to that of the characteristic decay
interface to conduction of a LTW. Therein the two interfaces, i.e., the former 
$+$ and $-$fronts, respectively, are in a robust
equilibrium with each other at a uniquely selected fixed distance $l$ that 
depends on $r,\psi$ as shown in Fig.~\ref{Fig:3Psisv+l+w-r}(b).

So the $-$interfaces of LTWs and front states do not influence significantly the
upstream part of these stuctures. Hence, the local concentration 'barrier'
ahead of the $-$interface does not select the drift velocity of LTWs as
speculated previously \cite{Barten95II}. It is rather the $+$interface that is
the more important one.

\subsection{Stabilization via front repulsion}

Note that it is the 'downstream wake' in the concentration field of the 
preceeding $+$front that
effectively slows down the $-$front: When the latter reaches the region where 
the mixing number $M(x)$ [Fig.~\ref{Fig:frontpics}(e)] starts to decrease towards
the preceeding $+$front, i.e., when the convective contribution 
$\langle b \rangle$ [Fig.~\ref{Fig:frontpics}(g)] to the local buoyancy starts to
increase then the speed of the approaching 
$-$interface has to slow down. This distance over which the $+$front influences 
the $-$front in a two-front structure grows when the Soret 
coupling becomes stronger. For example at $\psi>-0.3$ a separation of about 
160 rolls between the two interfaces is not sufficient to ensure independence. 

The sensitive dependence of the velocity of the $-$interface on
the concentration-induced buoyancy variation in the 'wake' behind the $+$front
is the main reason for the robust localization mechanism of (long) LTWs. The 
invasion of conduction into the convective region via the trailing $-$interface
is stopped at just that well defined distance from the $+$front where the 
concentration-induced convective buoyancy $\langle b \rangle$
has become sufficiently large. The latter increases monotonously towards the 
well mixed region under the $+$interface since this degree of advective mixing 
decreases gradually in the 'wake' behind the $+$interface. See, e.g., 
ref.~\cite{Jung02} for an explanation of the associated interplay of diffusion 
and advection which both reduce concentration gradients and the Soret effect 
which generates them. Of course, the effect of stopping the approaching 
$-$interface at a particular distance from the $+$interface can be interpreted
as an effective repulsive interaction between them.

\subsection{Long LTWs}

The structural similarity 
between long LTWs and fronts is documented in Fig.~\ref{Fig:frontpics}. 
Differences between the full lines (fronts) and the dashed ones (LTW) 
are visible only in the case of the $-$front in
Fig.~\ref{Fig:frontpics}(d,f,h). Here the bulk asymptotic TW that is realized 
to the left of the $-$front differs slightly from the plateau TW in the LTW. 

The above mentioned $\delta C$ redistribution via $<\vec{J}>$ enhances the
buoyancy at the $+$interface and leads there to a self-consistent 
stabilization of convection against invasion of conduction at the $+$front.
This mixing effect makes stable LTWs possible even for low heating rates $r$ 
where neither extended TW convection nor fronts exist. 

Long LTWs are characterized by a wide TW part with a well developed
plateau with almost no lateral variation in the convection properties like,
e.g., $v_p(x)$ or $M(x)$ \cite{Jung02}. The TW plateau separates the growth and 
the decay part of convection at the $+$ and $-$interface, respectively
and it provides a communication mechanism favoring one direction: The first 
region is shielded from the second one by the fast 'downstream' phase 
propagation. Like in a single $-$front state the $-$interface of the LTW does 
not influence the 'upstream' TW; it only manages the decay transition of the 
TW vortices into the quiesent fluid. Thus, the
$+$front character at the $+$interface is also present in the LTW. And the
properties of long LTWs are dominated by and similar to those of the single 
$+$front at the same $r$ if the latter exists. For example, 
the drift velocities of long LTWs agree with the values
of the corresponding $+$fronts in Figs.~\ref{Fig:Psi25v+om+k-r}, 
\ref{Fig:3Psisv+l+w-r}, \ref{Fig:4Psisom+r-k}, \ref{Fig:3DPsi40om-k-r}, 
\ref{Fig:LTWapproach}, \ref{Fig:4Psisom-r}. Furthermore, they 
continue to show the same linear variation with $r$ as the $+$fronts even where 
the latter cease to exist at smaller $r$, cf., the open
circles in Fig.~\ref{Fig:3Psisv+l+w-r}(a) for the cases of $\psi=-0.35$ and 
$\psi=-0.4$. Similarly, the variation $\omega(r)$ of long LTWs follows the
corresponding one of $+$fronts, cf., open circles and filled triangles in 
Fig.~\ref{Fig:4Psisom-r}. 

A comparison of LTW plateau values with extended TWs and front TWs 
is presented in Fig.~\ref{Fig:Psi25v+om+k-r}(b,c) for $\psi=-0.25$, in 
Fig.~\ref{Fig:3DPsi40om-k-r}  for  $\psi=-0.4$, and in 
Fig.~\ref{Fig:4Psisom+r-k} for for all examined $\psi$. At
$r^{LTW}_\infty=r^F_{eq}$ (arrows) 
there is no difference between the fronts, the diverging LTWs, and the extended
saddle TWs. 

For decreasing $r$ convection is less stable, the disintegration of the
traveling rolls sets in earlier, and the LTW length $l$ is therefore reduced,
cf., the inset of Fig.~\ref{Fig:Psi25v+om+k-r} and
Fig.~\ref{Fig:3Psisv+l+w-r}(b). The smallest plateau wave numbers of LTWs are 
realized for diverging lengths at $r^{LTW}_\infty(\psi)$. With decreasing $r$ 
one 
finds a slight increase of $k_{plateau}$ while the wave number selected by a 
single $+$front remains close to that of the saddle TW.

So this is the LTW bifurcation scenario that we found in the range 
$-0.4 \leq \psi \leq -0.25$ (for a discussion of the scenario at smaller Soret
coupling strength cf. Sec.~\ref{SEC:small-soret}): Approaching 
$r^{LTW}_\infty$ from below LTWs become 
indistinguishable from front states when $l \to \infty$. But further 
below $r^{LTW}_\infty$ LTWs
differ more and more in particular with respect to the bulk 
wave numbers as can be seen in
Figs.~\ref{Fig:4Psisom+r-k} and \ref{Fig:3DPsi40om-k-r}.
However, for long enough LTWs with a well developed spatial bulk plateau 
behavior the frequencies $\omega (r, \psi)$ and drift velocities $v_d(r, \psi)$ 
vary like the corresponding quantities of $+$fronts. This confirms the fact 
\cite{Jung02} that it is the $+$front-like growth 
interface that selects the properties of a long LTW. Shorter LTWs behave
also with respect to the variation of $\omega (r, \psi)$ and $v_d(r, \psi)$ 
somewhat differently.

\subsection{Short LTWs}

\begin{figure}
 \includegraphics[clip=true, angle=0,width=8.5cm]{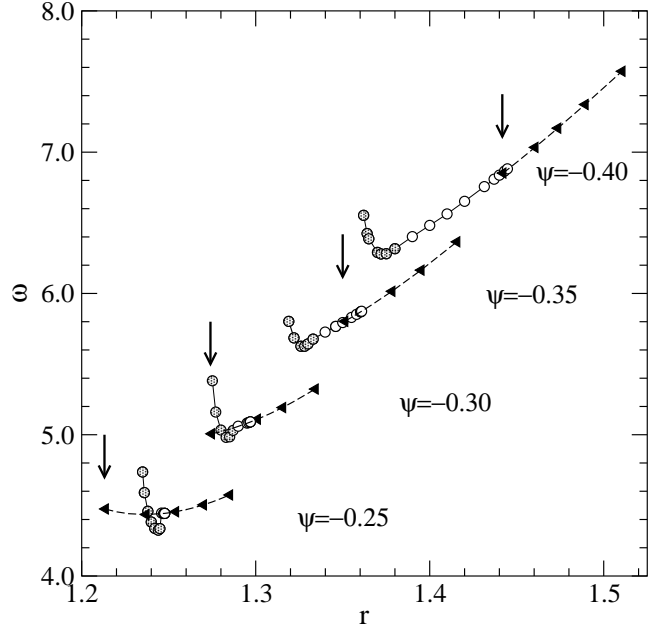}
 \caption{ Frequency $\omega$ of $+$fronts (dashed lines with triangles) and 
 of LTWs (full lines with circles) in the respective comoving frame versus 
 $r$ for different $\psi$. Open and shaded circles refer to long and short LTWs,
 respectively. The frequencies of the former are the same as those of the 
 fronts while short LTWs differ. Arrows indicate the lower limit of existence 
 of the fronts at $r^F_{min}=r^{TW}_s(k \simeq \pi)$. 
 \label{Fig:4Psisom-r}}
 \end{figure}

Reducing $r$ one eventually arrives for any $\psi$ at the regime of short LTWs
that are marked in Figs~\ref{Fig:Psi25v+om+k-r}, \ref{Fig:3Psisv+l+w-r}, and 
\ref{Fig:4Psisom-r} by shaded circles and that are located close to the dotted 
line in Fig.~\ref{Fig:phasediagram}. Here, the dominant influence of the $+$front
vanishes eventually in the regime of short LTW pulses.
No convection plateau can be identified any more in these structures, cf. the
dotted curves in Fig.~\ref{Fig:frontrelaxation}.
The prototype of a short LTW consists of a growth interface which is followed 
directly by the decay of convection so that the whole pulse has to be seen now 
as one integrated structure that no longer contains front-like independent $+$ 
and $-$interfaces. Hence, short LTWs show a strong lateral
variation of their properties. The shape of their amplitudes superficially 
resemble the pulse solutions of the complex Ginzburg-Landau equation 
\cite{Niemela90,Steinberg91,Kolodner91c,Thual8890,Malomed90,Hakim90,vSaarloos9092}. 

Like for long LTWs the stable existence of short pulses below any heating that 
is necessary to sustain 
extended TWs is caused by a lateral $\delta C$ redistribution over the pulse.
Also its frequency $\omega$ is constant in the frame that comoves with the drift
velocity $v_d$ of the pulse like for a long LTW. However, compared to long LTWs
short LTWs provide a qualitatively new convection structure.
They are independent of and cannot even be compared with extended TWs because 
there is no characteristic wavelength or phase velocity. The special 
character of short pulses compared with long LTWs or fronts is 
reflected in the change of the $r$-variation of $v_d, l, w_{max}, \omega$ that 
can be seen in Figs.~\ref{Fig:Psi25v+om+k-r}, \ref{Fig:3Psisv+l+w-r}, and 
\ref{Fig:4Psisom-r} by comparing the dashed circles with the open ones and the
filled triangles.

We observed the shortest possible LTWs at the lower end, $r^{LTW}_{min}$, of 
the $r$ band of LTWs. There they seem to end via a saddle 
node bifurcation for pulses \cite{Hakim90}. These minimal pulses always 
contained about 5 convection rolls 
for all Soret couplings $-0.65 \leq \psi \leq -0.08$ that we have 
investigated. This surprising universality of 
$l_{min} \simeq 5$, i.e., its insensitivity to the values of the actual heating
rate $r^{LTW}_{min}$ and the Soret coupling $\psi$ is still unexplained.  
 
Approaching the lower band limit of
LTWs their flow intensity steeply drops [Fig.~\ref{Fig:3Psisv+l+w-r}(c)]
and consequently their frequency increases [Fig.~\ref{Fig:4Psisom-r}] as the
degree of advective mixing of the fluid decreases.

\subsection{Comparison with LTW models} \label{SEC:LTWmodels}

Several attempts have been made to describe LTWs by simple model equations.
Stable pulse solutions of the complex Ginzburg-Landau equation (CGLE) were 
proposed 
as a model for confined binary mixture convection \cite{Thual8890}.
The nonlinear interaction between the local amplitude and frequency seems to be
the essential localization mechanism in this approximation.
Indeed, one could find localized solutions of increasing length up to
the limit of an infinitely long two-front state 
\cite{Malomed90,Hakim90,vSaarloos9092}.
But some basical problems remained: 
Within the CGLE all pulses drift with the same velocity. This is
the critical linear group velocity $v_g$ if the coefficients are derived from an 
asymptotic 
reduction of the full hydrodynamic field equations. But $v_g$ is too 
fast by a
factor of about 20-40 compared with the LTW drift velocity in experiments or 
simulations 
\cite{Cross88Knobloch88,Zimmermann8993,Surko87,Barten91,Luecke92,Barten95II,Jung02,Kolodner91a,Kolodner94}.
Brand and Deissler \cite{Brand8990} introduced asymmetry in the pulse properties
by adding nonlinear gradient terms to the CGLE. 
A similar extension was given by Bestehorn {\it et al.} \cite{Bestehorn89,Bestehorn888991}
within their framework of order parameter equations. Both could produce a very 
slow drift, even opposite to the phase direction \cite{Bestehorn90}.
But this kind of nonlinear modification of the linear group velocity involves a
balance which seems to be
too fragile to explain the occurrence of small pulse velocities over 
a whole range of $\psi$ and $r$ \cite{RieckeL92}.

Another problem was mentioned by van Saarloos and Hohenberg \cite{vSaarloos9092}.
According to their model of a quintic CGLE nonlinear wide pulses 
are expected by counting arguments to exist only in a codimension-2 
submanifold of 
the parameter space. Provided there are no hidden symmetries this seems to be 
incompatible 
with the robust occurrence of LTWs in experiments.
Furthermore, stable pulse solutions seem to exist only in the bistable regime 
whereas LTWs are known to
persist well above the linear onset of extended convection for weakly negative 
$\psi$ 
\cite{Niemela90,Behringer9091,Kolodner9091,Kolodner91a,Kolodner91c,Kolodner91d,Barten91,Barten95II,Luecke98}.
Furthermore, coexisting small stable and wide unstable LTWs were never seen in 
the CGLE but found in experiments by Kolodner \cite{Kolodner94}. Instead, 
stable broad pulse solutions
are found in the model to arise in a saddle node bifurcation together with an 
unstable branch
of smaller 'critical droplets' near the basic state \cite{Hakim90}.
Finally, numerical solutions of the field equations show the 
existence 
of stable LTWs even below the lowest TW saddle node \cite{Jung02}. This makes 
clear that LTWs are
influenced by a localization mechanism that is not contained in the CGLE.

Inspired by simulation results of Barten {\it et al.} \cite{Barten91,Luecke92} 
which showed the important role of the concentration field for a LTW Riecke 
\cite{RieckeL92,RieckeD92} 
proposed an extension of the Ginzburg-Landau equations. Within a weakly 
nonlinear expansion he coupled into the standard CGLE as an additional slow 
variable the amplitude ${\mathcal C}$ of an advected mean large-scale 
concentration mode that influences the growth of the critical modes.
A similar idea was advanced already by Glazier {\it et al.} \cite{Kolodner9091}.
The extension can induce an additional amplitude-intability of phase winding 
solutions 
to modulated waves. It may be considered as the origin for pulse formation in 
this ansatz \cite{Riecke01}. 
Riecke showed that the influence of the real ${\mathcal C}$-mode alone on 
the local growth rate (without dispersion) suffices
to generate slowly drifting stable pulse solutions even below a supercritical 
TW-bifurcation \cite{RieckeD92}.
In this way he modelled a new localization mechanism to explain the 
robust occurrence of LTWs in binary mixture convection.
The amplitude ${\mathcal C}$ can be interpreted as a measure for the local 
mixing state or the mean convection-induced deviation of the vertical
concentration gradient from the conductive one. 
In this way his extended complex Ginzburg-Landau equation (ECGLE) contains in a 
sketchy way physical effects like the mixing influence on the growth rate and 
the large-scale concentration redistribution. 

Riecke characterized within his model short and long LTWs 
as dispersion-dominated pulses \cite{Riecke96} and states of two fronts 
that are bounded by the ${\mathcal C}$ dynamics \cite{RieckeD95}, respectively.
He proposed an explanation for their coexistence in stable and unstable form, 
respectively,
by the competition between dispersion-dominated and ${\mathcal C}$-dominated 
localization \cite{RieckeD95,RieckeL95}.

Note, however, that
in contrast to the model used by Riecke our results show 
stable long LTWs that drift either in or opposite to
the direction of phase propagation depending on paramters. 
It would be interesting to check whether adding a term of the form 
$v |{\mathcal A}|^2 \partial_x {\mathcal C}$ to the ${\mathcal C}$-equation 
can stabilize forward drifting long pulses within the model since it 
models the concentration 'wake', i.e., the transport of the local mixing state 
in phase direction by the traveling rolls of amplitude ${\mathcal A}$.

Numerical and analytical investigations of the ECGLE predict a hysteretic 
transition from
slow to fast drifting pulses or the existence of oscillatory moving ones 
\cite{Riecke96}. But both were never seen in experiments or simulations.
Thus, despite their capability in elucidating some essential mechanisms 
CGLE type models have the drawback so far that they reproduce only 
single aspects of LTWs in a qualitative manner.
Their range of validity and their predictive power is not well known.
And since a satisfactory relation with the 
full field equations has not been established these models remain somewhat 
arbitrary. 

It appears questionable that weakly nonlinear expansions with spatially 
slowly varying mode amplitudes are approriate at all in view of the very 
large P\'eclet numbers, $\cal O$(1000), measuring the strength of the nonlinearity 
in the concentration balance. Thus, so far numerical simulations of the full 
field equations seem 
to be the appropriate tool besides careful experiments to gather insight into 
the specific physical mechanisms for LTW formation in binary mixture convection.

\section{Comparison with experiments and discussion} \label{SEC:Compare-exp}
\subsection{Small Soret coupling strength} \label{SEC:small-soret}

An inspection of Figs.~\ref{Fig:Psi25v+om+k-r}, \ref{Fig:3Psisv+l+w-r}, and
\ref{Fig:phasediagram}
shows that the lower band limit $r^{LTW}_{min}$ for the existence of short 
LTWs and the crossing value, $r^F_{eq}=r^{LTW}_{\infty}$, where the velocities 
of free fronts become equal approach each other when $|\psi|$ decreases. Thus, 
one can foresee an interval of moderately negative $\psi$ where
short and long LTWs can be found close to $r^{LTW}_\infty$. 
In this case the upper band limit for the existence of stable LTWs should be 
defined by a 
backward saddle-node bifurcation at $r_s^{LTW}$ where the branches of stable 
short and unstable 
long LTWs annihilate each other. 

\begin{figure}
 \includegraphics[clip=true, angle=0,width=8.5cm]{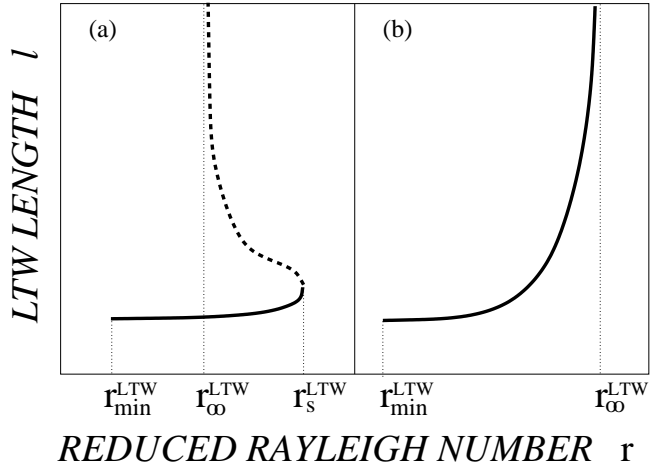}
 \caption{ Schematic bifurcation diagrams of LTW length $l$ versus $r$. (a)
  experimental results \cite{Kolodner94} obtained for $\psi=-0.127$ with dashed
  line denoting unstable states (cf., text for further explanation); (b) 
  numerically obtained bifurcation behavior for $-0.4 \leq \psi \leq -0.25$. 
 \label{Fig:l-r-scheme}}
 \end{figure}

Hence, we expect that the upper parts of the
bifurcation diagrams of $l$ versus $r$ in the inset of 
Figs.~\ref{Fig:Psi25v+om+k-r} and in Fig.~\ref{Fig:3Psisv+l+w-r}(b) curve
backwards towards smaller $r$ when $|\psi|$ decreases further below the values
of the two figures. In this way the shape of the curve $l(r)$ would change 
continuously from the form shown in Fig.~\ref{Fig:l-r-scheme}(b) to the one in 
Fig.~\ref{Fig:l-r-scheme}(a). The former shows schematically the bifurcation 
behavior of $l(r)$ that we have determined numerically for $\psi \lesssim -0.25$.
In fact, at $\psi$ slightly larger than -0.25 we expect the appearence of the 
saddle-node in the curves $l(r)$. Fig.~\ref{Fig:l-r-scheme}(a) is a schematic 
representation of experimental results of Kolodner \cite{Kolodner94} for 
$\psi=-0.127$ as presented in his Fig.~5. He stabilized by an adaptive 
heating mechanism long unstable LTWs in coexistence with short stable ones. 
Thus, the unstable LTW solution branch [dashed line in 
Fig.~\ref{Fig:l-r-scheme}(a)] forms for $r^{LTW}_\infty \leq r \leq r_s^{LTW}$ a 
separatrix between the domains of attraction of expanding two-front structures 
to the right of the dashed line in Fig.~\ref{Fig:l-r-scheme}(a) and the domain
to the left of the dashed line leading to stable narrow pulses or the basic 
state. Furthermore, small LTWs that are prepared at $r < r_s^{LTW}$  will 
evolve into expanding two-front structures when $r$ is increased above 
$r_s^{LTW}$.

Note that in Kolodner's experiment \cite{Kolodner94} done at $\psi=-0.127$ the 
upper band limit $r_s^{LTW}$ of LTWs lies {\it above} the Hopf bifurcation 
threshold $r_{osc}$ for extended TWs where perturbations of the quiescent fluid
can grow. Therefore, one has to address there questions related to linear 
and nonlinear 
convective versus absolute instability \cite{Chomaz92,Bu99}, to linearly selected
so-called pulled fronts versus nonlinearly selected so-called pushed fronts 
\cite{vSaarloos9092,vSaarloos03}, and to the robustness and stability of 
nonlinear fronts under emission or absorption of TW perturbations that can 
grow in the region occupied by the quiescent fluid.

\subsection{Strong Soret coupling strength}

For stronger negative $\psi$ the measured LTW properties agree qualitatively 
well with our results. For example, the 'arbitrary-width confined states' 
found in experiments \cite{Kolodner93II} for $\psi=-0.253$ 
at a single Rayleigh number are to be identified as two-front structures.
A quantitative comparison is difficult due to the difference in the boundary 
conditions:
We simulated two-dimensional convection assuming translational symmetry in 
$y$-direction while the narrow experimental convection channels
impose no-slip conditions at the walls perpendicular to the roll axes.
There are three effects that account for the difference beteen experiments and
simulations. 

First, the characteristic Rayleigh numbers
in the experiments are higher. This is already known from the 
suppression of oscillatory or steady convection instabilities in narrow channels
\cite{Platten84,CJT97,AlBa04}.
The no-slip conditions at the side-walls generate a nontrivial $y$-variation 
of the velocity 
field that introduces additional internal friction and that has to be 
compensated by a higher heating rate \cite{Catton72,Ohlsen90,Bensimon90}.

Second, the LTW drift velocities in the experiment have the global tendency to
lie below those of the simulations. For example, for $\psi=-0.253$ the
experimental LTWs \cite{Kolodner93II,Kolodner94} move opposite to the phase 
velocity whereas according to our calculations $v_d$ should be around $0.05$.
Again, we attribute this difference to the fact that we neglect gradients
in $y$-direction. They change the concentration redistribution dynamics in
particular at the $+$interface of the LTW which determines the drift velocity.
In this context one has to note that already weak
inhomogeneities in a convection cell can slow down and even pin the LTW 
movement \cite{Kolodner91a,Kolodner91c,Kolodner93II}.
Finally, the influence of the different boundaries on the frequencies, phase 
velocities, and wave numbers of confined states are totally unknown.
 
Also when comparing quantitatively our results for fronts with experiments one 
should take into account the above discussed points.

Extrapolating our results for the front velocities to more negative $\psi$
beyond $\psi=-0.4$ we see that already at the lowest TW saddle location 
$r_{min}^{TW}$ two-front stuctures
would expand with $v_F^- > v_F^+$. In other words, the velocity crossing 
point $r^F_{eq}$ is no longer above $r_{min}^{TW}$  but has virtually moved 
below
the lowest TW saddle location where in fact no fronts exist. On the other hand,
LTWs still exist in this $r$-range with length increasing with $r$. However, 
$l(r)$ does not seem to diverge anymore as for $-0.4 \lesssim \psi$ which is
compatible with the absence of fronts moving with the same velocity.

\subsection{Wall-attached confined structures}

Laterally confined convection patches of traveling rolls were found in the 
early experiments 
\cite{Moses87Heinrichs87,Fineberg88Steinberg89,Niemela90,Kolodner91b} 
that were done in narrow rectangular convection channels in the form of 
so-called 
wall-attached confined structures (WACS). They were localized near one of the 
short end walls closing the channel. 

These WACS can be understood with our 
knowledge of fronts and free LTWs. For example, for the weakly negative $\psi$ 
used in the early experiments the phase velocity of the WACS was directed
towards the wall to which they were attached. Indeed, for such parameters 
mainly short LTWs occur  
with  drift velocities in phase direction so that they would end as WACS of 
the above described type in finite length channels.
Furthermore, the measured WACS profiles of phase velocity $v_p(x)$, of 
wavelength $\lambda(x)$, and their decrease of frequency
with increasing $r$ \cite{Fineberg88Steinberg89,Steinberg91,Kolodner91b} 
agree qualitatively with the typical behavior of free short LTWs.

The connection between WACS and free LTWs was more explicitly demonstrated 
by Kolodner \cite{Kolodner90} for more negative separation ratios 
$\psi=-0.24$ and $-0.408$:
He prepared a free LTW pulse with large phase velocity (a 'fast confined state'
in his terminology) which drifted slowly opposite to the direction of
phase propagation towards an end wall of the convection channel and became 
there a WACS (a 'slow confined state' in his terminology) with lower phase 
velocity being directed away from the wall.
In this WACS the phase generating 'trailing front', i.e., the analogue of
the $+$interface is pinned at the wall and 
therefore without solutal gradients to the quiescent fluid as in a free LTW. 
The absence of these concentration variations at the $+$interface implies and 
allows a lateral concentration redistribution over the whole state at lower 
levels of the mixing number $M$ in WACS as compared to free LTWs. Consequently, 
the phase velocities and frequencies of WACS are smaller than those
of the respective free LTWs. A less dramatic drop of frequency was observed 
also between forward drifting LTWs and short WACS at $\psi=-0.047$ 
\cite{Steinberg91}.

Due to their better mixing capability, i.e., smaller $M$ it is 
very probable that short WACS 
can exist for heating rates below the lower band limit $r^{LTW}_{min}$ of 
stable LTW 
pulses --- at least in the case where the phase velocity is directed away 
from the wall. It would be very interesting to test this conjecture
experimentally, in 
particular for strongly negative $\psi$. There $r^{LTW}_{min}$ itself lies 
already 
well below the range of stable TWs \cite{Jung02} and so the WACS would lie even
lower.
An inportant hint that this conjecture is right is given by Ning {\it et al.} 
\cite{Ning96}. They have performed two-dimensional simulations 
of a finite-length convection channel
with realistic boundary conditions for $\psi=-0.47, \sigma=13.8$, and $L=0.01$. 
Neglecting the slight difference
in $\sigma$ their results should be comparable with our work.
They found short WACS at $r=1.35$ which is according to our results far below 
the TW saddle nodes for this
separation ratio and also below the lower band limit $r^{LTW}_{min}$ of free 
LTWs 
for $\psi=-0.40$. However, these authors claim --- we think, incorrectly ---
that their WACS lie above the saddle-node location $r_s^{TW}$ of extended TWs.

\subsection{LTW and front stability}
In the previous section we have shown that the $+$interface where the 
convection rolls grow in 'downstream' direction out of the quiescent fluid 
plays the dominant role for the stability of LTWs. While the $-$interface 
where the decaying rolls are advected into the quiescent fluid does 
not play a decisive role. This is clearly confirmed in 
pulse collision experiments \cite{Kolodner9091}.
 
Fast TW pulses --- linear ones with small amplitude as well as nonlinear 
ones with larger amplitude --- were completely absorbed by a LTW when the 
pulses hit the
$-$interface of the LTW, i.e., when the pulse velocity is directed opposite to 
and towards
the phase velocity of the LTW. Then the collision with the pulse affects only 
the $-$interface
itself and perturbations are quickly advected out of the LTW and do not propagate
upstream towards the $+$interface. For the same reason double-LTW states of 
two counter propagating waves can persist over a
long time \cite{Kolodner9091} or even be stable \cite{Kolodner91d}.
Moreover, a pair of LTWs that have their phase propagation directed towards 
each other and that interact with each other via their decay interfaces is seen 
to be stable over a substantially wider $r$-range than two LTW pulses which 
are connected at their growth interfaces \cite{Kolodner91d}. Obviously the 
latter case is more critical for the structural integrity of the involved LTWs.
 
A LTW is most likely destroyed when another wave with the same direction 
of phase propagation infiltrates its phaseflow at the growth region. 
Then, while growing the perturbations can be transmitted into the 
strongly nonlinear bulk part of the LTW and can destroy its coherence.

The different selection and stability properties of the $+$ and $-$interfaces
were already observed 
in transient convection behavior in various experiments (see for example 
\cite{Kolodner88,Bensimon90}), however without further investigation.
 
\subsection{Defected confined states}

The fact that ($i$) different $-$fronts with different bulk 
TW parts are possible as stable coexisting states
and that ($ii$) a $+$front interface is in general stable against
downflow perturbations opens 
the possibility for another kind of stable long 
confined TWs: Therein rolls with low wavelength grow out of the quiescent 
fluid in a 'normal' growth part. In the bulk an incoherent phase front 
connects this fast wave that is coming from the $+$interface with a slow wave
of higher wavelength and larger amplitude
via spatiotemporal dislocations as, e.g., in Fig.~\ref{Fig:irregularfront}.
The transition could take place via one or more intermediate convection 
states. Eventually this slow TW convection undergoes a decay transition
into the basic state via a coherent $-$interface. Such 'defected 
confined states' are indeed observed in annular containers \cite{Bensimon90,
Behringer9091} and were studied by Kolodner \cite{Kolodner94}. 
He found such structures only for $\psi\le -0.21$.
One may speculate that for these separation ratios the bulk TW that is selected 
by and behind the $+$interface is absolutely unstable against the slow wave 
in the further 'downstream' part.
This could explain the existence of persistent roll pair annihilations without 
the need of fluctuations.

We finally mention that the occurrence of phase annihilating dislocations 
between a growth part and the downstream convection was observed for moderately 
negative $\psi$ already in narrow rectangular containers \cite{Kolodner91b}.
Furthermore, end-wall induced Eckhaus instabilities of downstream TW states 
have been seen in simulations \cite{BuLu99}. 

\section{Conclusion}

 For parameters where the conductive quiescent fluid is stable and where 
 spatially extended TW solutions bifurcate subcritically out of it we have 
 investigated in quantitative detail relaxed, strongly nonlinear oscillatory 
 convection 
 structures with one or two interfaces to the quiescent fluid, i.e, fronts and
 LTWs, respectively. They are time-periodic global
 nonlinear modes: in the frame that is comoving 
 with the respective front velocity $v_F$ or with the LTW drift velocity $v_d$
 the oscillations have everywhere the same period. 
 
 Fronts come in two varieties.
In a $+$front state the quiescent fluid is located
'upstream', i.e., phase propagates out of it into convection. 
In a $-$front the quiescent fluid is located
in 'downstream' direction and phase moves out of convection 
into conduction.

The lowest Rayleigh number for the existence of fronts is 
the lowest saddle-node location of extended 
TWs, $r^F_{min} = r^{TW}_{min}$: below it there are no TWs to which the 
interface from conduction can 
connect. However, LTWs of {\em finite} length $l$ can coexist bistably
together with the conductive state well below the lowest TW saddle when the
Soret coupling is sufficiently negative, $r^{LTW}_{min} < r^{TW}_{min}$. 
Furthermore, we have arguments 
that WACS at end walls of rectangular channels can exist even at smaller $r$ 
than LTWs. 

Central for understanding fronts and LTWs is a large-scale concentration
redistribution that influences the buoyancy at the interfaces to conduction in
different ways than in the bulk TW parts. For example, at the $+$interfaces of
fronts and LTWs alike there is a buoyancy overshoot which is sufficiently 
large to sustain local convection growth there and that can cause even 
invasion of convection into the stable quiescent fluid. At the $-$interface
the lateral buoyancy variation is such as to induce the decay of the approaching
convection rolls into the conductive state. 

Front velocities as well as LTW drift velocities are much smaller than the phase
velocities of the carrier waves for reasons that are related to the 
concentration redistribution dynamics. The velocities of $+$fronts decrease 
with growing $r$ while those of $-$fronts increase. At some
$r^F_{eq}$ they become equal so that both fronts move with the same velocity.
At this Rayleigh number the length $l$ of the LTWs diverges and there, and 
strictly speaking only there, the limiting LTW can be seen as a state 
consisting of two fronts. However, $+$fronts and long LTWs have almost 
identical propagation velocities and frequencies. Furthermore, they select a 
similar bulk wave number. The selected frequencies and bulk wave numbers are
close to those of a saddle-node TW. In fact, it is the $+$front-like growth 
interface that selects the properties of long LTWs. 

Small amplitude extended TW perturbations of the conductive state oscillate 
with the large Hopf frequency. But the global-mode oscillation 
is restrained by the requirement that its frequency has to allow stable 
developed bulk TW convection. It is interesting to note that the $+$interface 
connecting conduction with
convection selects the largest possible frequency eigenvalue that meets this
requirement, namely the TW saddle-node frequency.
All our $+$fronts select bulk TW wave numbers close to the large-$k$ 
branch of the TW saddle-node curve, i.e., wave numbers that are too large 
to be Eckhaus stable. However, these TWs are only convectively unstable: 
perturbations can grow but while doing so they
are advected sufficiently fast downstream in the direction of the TW phase 
propagation so that they cannot influence the upstream part of the $+$front 
state in a persistent way. 

While $+$front states seem to be uniquely selected we could produce for a 
fixed $r$ 
different coherent $-$fronts that were characterized in the bulk part
by different wave numbers and frequencies close to the TW saddles.
The  decay interface adjusts itself to the respective bulk TW part but does not
exert an influence in 'upstream' direction on the bulk convection within a coherent
$-$front. In contrast, the growth under a $+$interface 
induces in downstream direction a long concentration 'wake' that is 
characteristic
for $+$fronts and long LTWs and of special importance for the latter.

Here it is interesting to notice that all the interfaces of fronts and LTWs
consist typically only of about 3-4 convection rolls. We furthermore should
like to mention that $+$interfaces of fronts and LTWs always locate a minimal 
wavelength. Its value, $\lambda_{min}\sim 1.4$, is remarkably universal for
{\it all $r$ and $\psi$} that we have simulated. This is unexplained so far. 

We have also prepared initial two-front stuctures by connecting a $+$front and 
a $-$front with a common long bulk TW. When $r>r^F_{eq}$ they expand. But at 
$r<r^F_{eq}$ they shrink towards a uniquely selected LTW of fixed length $l$.
Here the 'downstream wake' in the concentration field of the 
preceeding $+$front exerts an effective repulsion on the approaching $-$interface: 
the invasion of conduction via the latter is stopped at a well
defined distance $l$ that is determined by the concentration-induced buoyancy 
levels in the 'wake' of the $+$front. 

LTWs shortly below $r^F_{eq}$ (where LTW with diverging $l$ are possible) are 
very long. Their drift velocities, frequencies, and many stuctural properties 
are similar to those of $+$fronts. Decreasing $r$ the LTW length decreases
and one eventually arrives for any $\psi$ at the regime of short LTWs that lies
for strongly negative $\psi$ well below the TW saddle-nodes. These short LTWs
without a convection plateau are qualitatively different structures.
This is also reflected by their drift velocities and frequencies
showing a variation with $r$ that differs from those of long LTWs. 
The shortest possible LTWs are realized at the lower end of 
the $r$ band of LTWs. These minimal pulses always 
contained about 5 convection rolls for all Soret couplings that we have 
investigated. This surprising universality of $l_{min} \simeq 5$ remains to be 
explained.  

\clearpage

 \clearpage


 \end{document}